\documentclass[a4paper,noarxiv,unpublished,twocolumn]{quantumarticle}
\pdfoutput=1

\usepackage[utf8]{inputenc}
\usepackage[english]{babel}
\usepackage[T1]{fontenc}
\usepackage{amsmath}
\usepackage{hyperref}
\usepackage{tikz}
\usepackage{lipsum}

\usepackage[numbers,sort&compress]{natbib}

\usepackage{mathtools}
\usepackage{amssymb}
\usepackage{braket}

\usepackage{color}
\usepackage{bm}
\usepackage{xcolor}
\usepackage{graphicx}
\usepackage{psfrag}
\usepackage[normalem]{ulem}
\usepackage{siunitx}
\usepackage{amsfonts}

\usepackage{qcircuit}
\usepackage[linesnumbered,ruled,vlined]{algorithm2e}
\usepackage{textcomp}
\usepackage{wasysym}

\usepackage{amsthm}

\newtheorem*{theorem-non1}{Weighted Sum}
\newtheorem*{theorem-non2}{Controlled Weighted Sum}
\newtheorem*{theorem-non3}{Sum of Products}
\newtheorem*{application1}{Expected Value}
\newtheorem*{application2}{Restricted Weighted Sum}
\newtheorem*{application3}{Mean Value}

\emergencystretch=\maxdimen
\DeclarePairedDelimiter{\innerprod}\langle\rangle

\begin{document}

%\preprint{}
\title{A Generalized Quantum Inner Product and Applications to Financial Engineering}% Force line breaks with \\
%\thanks{}%

\author{Vanio Markov}
\affiliation{Wells Fargo}
\author{Charlee Stefanski}
\affiliation{Wells Fargo}
\affiliation{University of California, Berkeley}
\author{Abhijit Rao}
\author{Constantin Gonciulea}
\orcid{0000-0001-5870-4586}
\affiliation{Wells Fargo}

\maketitle
\begin{abstract}
    In this paper we present a canonical quantum computing method to estimate the weighted sum
    $\sum_{k=0}^{2^n-1} w_k f(k)$ of the values taken by a discrete function $f: \{0, \mathellipsis, 2^n-1\}
    \rightarrow \{0, \mathellipsis, 2^m-1\}$ for $n, m$ positive integers and weights $w_k \in \mathbb{R}$ for $k
    \in \{0, \mathellipsis, 2^n-1\}$. The canonical aspect of the method comes from relying on a single
    linear function encoded in the amplitudes of a quantum state, and using register
    entangling to encode the function $f$.

    We further expand this framework by mapping function values to hashes in order to
    estimate weighted sums of hashed function values $\sum_{k=0}^{2^n-1} w_k h_{f(k)}$,
    with $h_v \in \mathbb{R}$ for $v \in \{0, \mathellipsis, 2^m-1\}$. This generalization
    allows the computation of restricted weighted sums such as value at risk, comparators,
    as well as Lebesgue integrals and partial moments of statistical distributions.

    We also introduce essential building blocks such as efficient encodings of standardized
    linear quantum states and normal distributions.

\end{abstract}

\maketitle

%%%%%%%%%%%%%%%%%%%%%%%%%%%%%%%%%%%%%%%%%%%%%%%%%%%%%%%%%%%%%%%%%%%%%%%%%%%
\section{\label{sec:introduction}Introduction}
%%%%%%%%%%%%%%%%%%%%%%%%%%%%%%%%%%%%%%%%%%%%%%%%%%%%%%%%%%%%%%%%%%%%%%%%%%%

Key areas of Financial Engineering such as Risk Management, Portfolios Optimization, Financial Data Analysis,
and Machine Learning use methods of Linear Algebra and Fourier Analysis. A key operation in these areas is the inner 
product of two vectors - the sum of the products of their 'inner' corresponding components. Since the financial
analysis is usually performed in very large vector spaces the efficient implementation of the inner product is
critical for the feasibility of the Quantitative Financial Technology.

If a financial problem is modeled through the state space of a quantum system, then the inner product leverages quantum
parallelism of quantum transformations. Quantum parallelism, together with interference, allows quantum algorithms
such as Grover's search~\cite{Grover1996Search} and quantum Fourier transform~\cite{Coppersmith1994AnAF} to achieve
proven speedup compared to their classical counterparts.

In this paper we make the following contributions:
\begin{enumerate}
    \item A generalized inner product that uses both quantum state and amplitude encoding
    \item Standardized building blocks as a unifying framework for calculating weighted sums
    \item Applications to weighted sum computations that occur in financial engineering, e.g. expected values, value
    at risk (VaR), derivative pricing, etc.
    \item A canonical method for computing the expected value of discrete functions
    \item Efficient methods for approximate and exact encodings of normal distributions
    \item Efficient methods for approximate and exact encodings of linear functions, including a canonical identity function used
    in expected value computations
\end{enumerate}

The canonical approach for computing quantum inner products leverages our prior work on encoding a binary polynomial
representing a discrete function ~\cite{gilliam2021foundational, CPBO}, by adding weights to both its inputs and
outputs. In this paper we will refer to polynomials of binary variables as binary polynomials.

In some of the inner product applications, we use circuits that efficiently encode linear
and rational functions, and normal distributions for a given number of qubits.

\bigskip
The paper is organized as follows:

Section~\ref{sec:prelim} introduces mathematical concepts and notation used throughout the paper, including
details about encoding binary polynomials representing discrete functions.

Section~\ref{sec:methods} presents the methods we propose for computing inner products.

Section~\ref{sec:distributions} discusses the preparation of quantum states representing standard distributions and
functions. In particular, we propose methods for exact or approximate encoding of the raised cosine and the normal
distributions, as well as linear functions.

Section~\ref{sec:applications} presents applications of the proposed framework.

Section~\ref{sec:experiments} presents results of execution of some of the algorithms on IBM Q
hardware~\cite{IBMQServices}.

Section~\ref{sec:related_work} includes references to similar work.

Section~\ref{sec:conclusions} contains concluding remarks.

Appendix~\ref{sec:catalog} contains a catalog of efficient circuits for
exact encoding of linear quantum states and normal distributions.

%%%%%%%%%%%%%%%%%%%%%%%%%%%%%%%%%%%%%%%%%%%%%%%%%%%%%%%%%%%%%%%%%%%%%%%%%%%
\section{\label{sec:prelim}Preliminaries}
%%%%%%%%%%%%%%%%%%%%%%%%%%%%%%%%%%%%%%%%%%%%%%%%%%%%%%%%%%%%%%%%%%%%%%%%%%%

This section contains some mathematical background necessary to understand the main results of this paper.

\subsection{\label{subsec:function_poly}Discrete Functions As Binary Polynomials}

As stated in~\cite{booleanfunctions}, any function $f: \{0, \mathellipsis, 2^n-1\} \rightarrow \mathbb{C}$, for a
given integer $n >0$, can be expressed as a polynomial $p: \{0, 1\}^n \rightarrow \mathbb{C}$:

\begin{equation*}
    p(x_0, \mathellipsis, x_{n-1}) = \sum_{k=0}^{2^n-1} f(k) p_k(x_0, \mathellipsis, x_{n-1})
\end{equation*}

where  $p_k$ for $0 \le k < 2^n$ are polynomials that satisfy the property

\begin{equation*}
    p_k(x_0, \mathellipsis, x_{n-1}) =
    \begin{cases}
        1, & \text{if } x_j = k_j \text{ for all } 0 \le j < n \\
        0, & \text{otherwise} \\
    \end{cases} \\
\end{equation*}

with $k = \sum_{j=0}^{n-1}k_j 2^j$ being the binary expansion of $k$, with binary digits $k_j \in \{0, 1\}$, for $0 \le
j < n$.

Therefore,

\begin{equation*}
    f(k) = p(k_0, \mathellipsis, k_{n-1}).
\end{equation*}

The polynomials $p_k$ for $0 \le k < 2^n - 1$ can be defined by

\begin{equation}
    \begin{split}
        p_k(x_0, \mathellipsis, x_{n-1}) & =  \left(\prod_{\substack{j \\ k_j = 1}} x_j \right) \left
        (\prod_{\substack{j \\ k_j = 0}} ( 1- x_j )\right) \\
        & = \prod_{j=0}^{n-1} \left((2k_j - 1)x_j  + 1 - k_j\right)
    \end{split}
\end{equation}

Recall that the square of a binary value $x \in \{0,1\}$ is the value itself ($x^2 = x$), because $0^2 = 0$ and $1^2
= 1$.

\subsection{\label{subsec:geometric_state}Geometric Sequence State}

For a quantum register with $m>0$ qubits and an angle $\theta \in \mathbb{R}$ let us define the state

\begin{equation}
    \label{eqn:geom_state_eqn}
    \begin{split}
        \ket{\gamma_{\theta}} & = \frac{1}{\sqrt{M}}\sum_{k=0}^{M-1} e^{i k\theta} \ket{k}_m \\
        & = \frac{1}{\sqrt{M}}\sum_{k=0}^{M-1}\left(\cos(k\theta) + i\sin(k\theta) \right) \ket{k}_m
    \end{split}
\end{equation}

where $M = 2^m$. Note that the amplitudes of this quantum state form a geometric sequence. This is one of the
building blocks--referred to as a complex exponential signal--in Digital Signal Processing. In quantum computing this
state plays an important role in the phase estimation and other algorithms, where the period of the signal
represented by the amplitudes encodes some relevant information.

\begin{figure}[ht]
    \centering
    \mbox{
        \Qcircuit @C=1em @R=0em @!R {
            0 & {} & {} & {} & \gate{P(2^{m - 1}\theta)} & \qw & \qw  \\
            \vdots & & & & \ldots \\
            m-1-i & {} & {} & {} & \gate{P(2^{i}\theta)} & \qw & \qw  \\
            \vdots  & & & & \ldots \\
            m - 1 & {} & {} & {} & \gate{P(\theta)} & \qw & \qw   \\
        }
    }
    \caption{The quantum circuit for encoding the state represented in Eq.~\ref{eqn:geom_state_eqn} in an $m$-qubit
    register. The circuit performs a series of Phase gates, denoted by $P$, using multiples of a given angle $\theta
    \in \mathbb{R}$.}
    \label{fig:geom_circuit}
\end{figure}
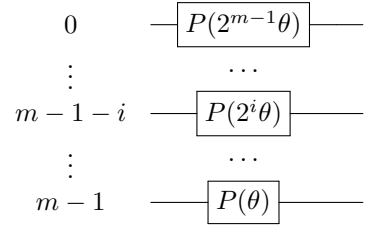

This state can be prepared with the circuit in Fig.~\ref{fig:geom_circuit}, where $P$ is the phase single qubit gate.

\subsection{\label{subsec:value_encoding}Integer Encoding}

Given a quantum register with $m>0$ qubits and $M = 2^m$. To encode a given integer value $-\frac{M}{2} \le
j < \frac{M}{2}$, we first create the state $\ket{\gamma_{j \frac{2\pi}{M}}}$, as described in
Section~\ref{subsec:geometric_state}, and then apply the inverse Fourier transform operator ($QFT^\dagger$) to this
state, resulting in the state $\ket{j}$ if $j \ge 0$ or $\ket{j+M}$ if $j < 0$, representing the Two's Complement
representation of $j$, as described in ~\cite{gilliam2021foundational, CPBO}.

\subsection{\label{subsec:binary_poly_encoding}Binary Polynomial Encoding}

Assume we have two quantum registers, a key register $\ket{k}_n$ with $n > 0$ qubits, and a value register
$\ket{v}_m$ with $m > 0$ qubits. We'll use the notations $N=2^n$ and $M=2^m$.

For a binary polynomial $p: \{0, 1\}^n \rightarrow \{0, 1\}^m$ with integer coefficients we can use the Quantum
Dictionary pattern, as introduced in ~\cite{gilliam2021foundational, CPBO}, to encode key-value pairs $(k, p(k))$,
with $0 \le k < N$ and $-\frac{M}{2} \le p(k) < \frac{M}{2}$, in the key and value registers, respectively, by using a
controlled version of the value encoding procedure described in Section~\ref{subsec:value_encoding}.

The registers are entangled in such a way that when measured, the outcome will consist of a key and its
corresponding value. The key register $\ket{k}_n$ represents the inputs of the polynomial $p$, and the value register
$\ket{v}_m$ represents the outputs.

Here is a brief overview of the implementation. We start by putting both registers in equal superposition:

\begin{equation}
    \label{eqn:equal_superposition}
    \frac{1}{\sqrt{N}}\sum_{k=0}^{N-1} \ket{k}_n \frac{1}{\sqrt{M}}\sum_{v=0}^{M-1} \ket{v}_m
\end{equation}

Any binary polynomial $p$ of $n$ variables with integer coefficients can be expressed as a sum of monomials:

\begin{equation}
    \label{eqn:sum_monomials}
    p(x_0, \mathellipsis, x_{n-1}) = \sum_{J \subseteq \{ 0, \mathellipsis, n-1 \}} c_J \prod_{j \in J}x_j \\
\end{equation}

For a monomial $c_J \prod_{j \in J}x_j$ we encode the integer $c_J$ using the value encoding procedure controlled on
the qubits corresponding to $J$ ~\cite{CPBO}. The resulting state is

\begin{equation}
    \label{eqn: mon_encoding_state_before_iqft}
    \frac{1}{\sqrt{N}}\sum_{k=0}^{N-1} \ket{k}_n \ket{\gamma_{p(k) \frac{2\pi}{M}}}_m.
\end{equation}

Finally, we apply the inverse Fourier transform operator to the value register and end up with the state

\begin{equation}
    \label{eqn: mon_encoding_state_after_iqft}
    \frac{1}{\sqrt{N}}\sum_{k=0}^{N-1} \ket{k}_n \ket{p(k)}_m.
\end{equation}

%%%%%%%%%%%%%%%%%%%%%%%%%%%%%%%%%%%%%%%%%%%%%%%%%%%%%%%%%%%%%%%%%%%%%%%%%%%
\section{\label{sec:methods}Inner Product Computation Methods: Simple and Generalized}
%%%%%%%%%%%%%%%%%%%%%%%%%%%%%%%%%%%%%%%%%%%%%%%%%%%%%%%%%%%%%%%%%%%%%%%%%%%

Let $V$ be the vector space describing an $n$-qubit quantum system. The inner product of two quantum states,
$\ket{\psi_A}$ and $\ket{\psi_B}$, is defined as

\begin{equation}
    \label{eqn:inner_prod_def}
    \innerprod{\psi_B|\psi_A } = \sum_{i = 0}^{N-1} \overline b_{i}a_{i}
\end{equation}

where $N = 2^n$ and the states $\ket{\psi_A}$ and $\ket{\psi_B}$ are prepared by the unitary transformations $A$ and $B$:

\begin{equation}
    \label{eqn:transformations_a_and_b}
    \begin{split}
    \ket{\psi_A}=A \ket{0}_n & = \sum_{i = 0}^{N-1} a_{i}\ket{i}_n \\
    \ket{\psi_B}=B \ket{0}_n & = \sum_{i = 0}^{N-1} b_{i}\ket{i}_n
    \end{split}
\end{equation}

\subsection{\label{subsec:inner_product}Inner Product of Quantum States Using State Preparation Operators}

The inner product of quantum states is invariant to unitary transformations, therefore we can apply the state
preparation unitary $B$ and represent $\innerprod{\psi_B|\psi_A}$ as

\begin{equation}
    \label{eqn:state_prep_unitary_b}
    \innerprod{\psi_B|\psi_A } = \innerprod{\psi_B|BB^{\dagger}|\psi_A }
\end{equation}

From Equation~\ref{eqn:transformations_a_and_b} we have

\begin{equation*}
        \bra{\psi_B}B =\bra{0}_n,
\end{equation*}

and therefore the inner product can be expressed as

\begin{equation*}
    \langle\psi_B | \psi_A\rangle = \bra{0} B^{\dagger}A  \ket{0}.
\end{equation*}

\begin{figure}[ht]
    \centering
    \mbox{
        \Qcircuit @C=1em @R=1em {
            & \lstick{\ket{q}_0{}}      & \qw    & \qw     & \multigate{4}{A}   &\multigate{4}{B^{\dagger}}    & \qw\\
            &                                      \\
            &              & \vdots                       \\
            &                                    \\
            & \lstick{\ket{q}_{n-1}{}}  & \qw    & \qw     & \ghost{A}          &\ghost{B^{\dagger}}           & \qw\\
        }
    }
    \caption{The quantum circuit that computes the inner product represented in Eq.~\ref{eqn:inner_prod_def} using
    the unitary operators $A$ and $B^\dagger$, as defined in Eq.~\ref{eqn:transformations_a_and_b}.}
    \label{fig:dagger}
\end{figure}
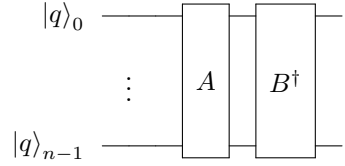

If the quantum state created by the circuit in Fig.~\ref{fig:dagger} is

\begin{equation*}
    \label{eqn:b_dag}
    B^{\dagger}A\ket{0}_n =  \gamma_0\ket{0}_n + \sum_{i = 1}^{N-1} \gamma_{i}\ket{i}_n
\end{equation*}

then the amplitude of the state $\ket{0}_n$ is the inner product we want to calculate:

\begin{equation}
    \label{eqn:state_0_inner_prod}
    \langle\psi_B | \psi_A\rangle=\gamma_0.
\end{equation}

Let us consider the following problem context and solution for the computation of weighted sums using inner product:

\begin{theorem-non1}
    \label{pattern1}
    Given an integer $n > 0$ and $N = 2^n$, weights $w_k \in \mathbb{R}$ defined for integers $0 \le k < N$ and a
    function $f: \{0, \mathellipsis, N-1\} \rightarrow \mathbb{R}$ we are interested in calculating the weighted sum

    \begin{equation*}
        \label{eqn:weighted_sum}
        \sum_{k=0}^{N-1}w_k f(k).
    \end{equation*}
\end{theorem-non1}

\begin{proof}[Solution]
    Let $A$ be an operator that prepares a state $\sum_{k =	 0}^{N-1}a_k\ket{k}_n$, with $a_k = a w_k$, for $0 \le k <
    N$, where $a \in \mathbb{R}$ is a common factor, and let $B$ be an operator that prepares the state $\sum_{k =
    0}^{N-1}b_k\ket{k}_n$ with $b_k = b f(k)$, for $0 \le k < N$, where $b$ is a common factor.

    Using the circuit represented in Fig.~\ref{fig:dagger}, we can compute the inner product
    $\sum_{k=0}^{N-1}a_k b_k$ as

    \begin{equation*}
        \label{eqn:inner_prod_weights}
        E_{\ket{0}} \coloneqq \bra{0}B^{\dagger}A  \ket{0}_{n}  = \sum_{k=0}^{N-1}a_k b_k = a b \sum_{k=0}^{N-1}w_k f
        (k),
    \end{equation*}

    therefore,

    \begin{equation*}
        \label{eqn:weights_sum}
        \sum_{k=0}^{N-1}w_k f(k) = \frac{1}{a b} E_{\ket{0}}
    \end{equation*}

    where $E_{\ket{0}} = \bra{0}B^{\dagger}A \ket{0}_{n}$ is the amplitude of $\ket{0}_n$ at the end of the computation.

    The desired weighted sum is the amplitude of the initial state $\ket{0}_{n}$ after applying the unitary operator
    $B^{\dagger}A$ to it, which can be estimated using amplitude estimation algorithms.
\end{proof}

\subsection{\label{subsec:generalized_inner_product}Generalized Inner Product}

Let us consider two quantum registers, key and value, with $n$ and $m$ qubits and $N=2^n, M=2^m$ computational states,
respectively. Given a function $f: \{0, \mathellipsis, N-1\} \rightarrow \{0, \mathellipsis, M-1\}$, we entangle the
two registers using the quantum dictionary pattern discussed in Section~\ref{subsec:binary_poly_encoding}.

Let $\ket{\psi_A}_n$ be the state of the key register prepared by the unitary operator $A$:

\begin{equation*}
    \label{eqn: unitary_a}
    \ket{\psi_A} = A\ket{0}_{n} = \sum_{k = 0}^{N-1}a_k\ket{k}_n.
\end{equation*}

The state of the two registers is

\begin{equation*}
    \label{eqn:state_after_op_a}
    (A \otimes I_m) \ket{0}_{n+m} = \ket{\psi_A} \ket{0}_m  = \sum_{k = 0}^{N-1}a_k\ket{k}_n\ket{0}_m.
\end{equation*}

The unitary operator $F$ entangles the two registers leading to the state

\begin{equation}
    \label{eqn:state_after_f}
    \begin{split}
        F (A \otimes I_m)\ket{0}_{n+m} & = F\left(\sum_{k = 0}^{N-1}a_k\ket{k}_n\ket{0}_m\right) \\
        & = \sum_{k = 0}^{N-1}a_k\ket{k}_n\ket{f(k)}_m.
    \end{split}
\end{equation}

Let $\ket{\psi_B}$ be the state of the value register prepared by the unitary operator $B$:

\begin{equation*}
    \label{eqn:unitary_b}
    \ket{\psi_B} = B\ket{0}_m = \sum_{v = 0}^{M-1}b_v\ket{v}_m.
\end{equation*}

When applying Hadamard gates to the qubits in the key register, we obtain the following state for the two registers:

\begin{equation}
    \label{eqn:h_and_unitary_b}
    (H^{\otimes n} \otimes B)\ket{0}_{n+m} = \frac{1}{\sqrt{N}}\sum_{\substack{0 \le k < N \\ 0 \le v < M}}
    b_v\ket{k}_n\ket{v}_m
\end{equation}

Applying $(H^{\otimes n} \otimes B^{\dagger})$ to the state represented in Equation~\ref{eqn:state_after_f} results in

\begin{equation*}
    \label{eqn:state_after_a_f_b}
    (H^{\otimes n} \otimes B^{\dagger}) F (A \otimes I_m)\ket{0}_{n+m}.
\end{equation*}

Therefore,

\begin{equation}
    \label{eqn:inner_prod_f_b}
    \bra{0}_{n+m} (H^{\otimes n} \otimes B^{\dagger}) F (A \otimes I_m)\ket{0}_{n+m} = \sum_{k = 0}^{N - 1}a_k b_{f(k)}
\end{equation}

The inner product of the amplitudes of the states in Equations~\ref{eqn:state_after_f} and~\ref{eqn:h_and_unitary_b}
is the amplitude of the computational state $\ket{0}_{n+m}$.
We refer to the right hand side of Equation~\ref{eqn:inner_prod_f_b} as a generalized inner product.

\begin{figure}[ht]
    \centering
    \mbox{
        \Qcircuit @C=1em @R=1em {
            & \lstick{\ket{v}_0{}}      	& \qw & \qw   				& \qw & \multigate{5}{F}    & \qw   &\multigate{2}{B^{\dagger}}    &\qw   \\ % key 0
            & \lstick{\textcolor{white}{0}} & \vdots \\
            & \lstick{\ket{v}_{m-1}}      	& \qw & \qw      			& \qw & \ghost{F}           & \qw   &\ghost{B^{\dagger}}           & \qw   \\ % key n
            & \lstick{\ket{k}_0{}}      	& \qw & \multigate{2}{A}    & \qw & \ghost{F}           & \qw   & \multigate{2}{H} & \qw   \\ % value 0
            & \lstick{\textcolor{white}{0}} & \vdots \\
            & \lstick{\ket{k}_{n-1}}      	& \qw & \ghost{A}           & \qw & \ghost{F}           & \qw   & \ghost{H} & \qw   \\ % value m
        }
    }
    \caption{The quantum circuit for computing generalized inner products. The operator $A$ encodes a given
    distribution of weights on the key register $\ket{k}_n$, and the operator $F$ entangles the inputs and outputs
    of a function. Applying $B\dagger$ to the value register $\ket{v}_m$ and $H^{\otimes n}$ to the key register
        $\ket{k}_n$, results in a state which contains the desired inner product in the amplitude $\ket{0}_{n+m}$.}
    \label{fig:dict_with_dagger}
\end{figure}
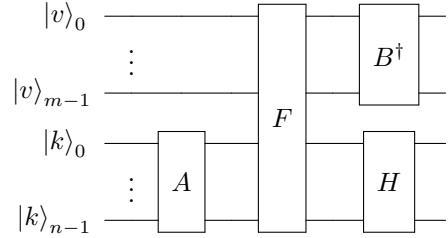

Let us consider the following problem context and solution for the computation of weighted sums using generalized
inner products:

\begin{theorem-non2}
    \label{pattern2}
    Given integers $n, m > 0$ and $N = 2^n, M = 2^m$, weights $w_k \in \mathbb{R}$ defined for integers $0 \le k < N$,
    weights/hashes $h_v \in \mathbb{R}$ defined for integers $0 \le v < M$ and a function $f: \{0, \mathellipsis,
    N-1\} \rightarrow \{0, \mathellipsis, M-1\}$, we are interested in calculating the weighted sum of
    weighted/hashed function values,

    \begin{equation*}
        \sum_{k=0}^{N-1}w_k h_{f(k)}.
    \end{equation*}
\end{theorem-non2}

\begin{proof}[Solution]
    Let $A$ be an operator that prepares a state $\sum_{k = 0}^{N-1}a_k\ket{k}_n$, with $a_k = a w_k$, for $0 \le k <
    N$, where $a \in \mathbb{R}$ is a common factor, and let $B$ be an operator that prepares the state $\sum_{v =
    0}^{M-1}b_v\ket{v}_m$ with $b_v = b h(v)$, for $0 \le v < M$, where $b \in \mathbb{R}$ is a common factor, and
    $F$ an operator that encodes the function $f$.

    Using the procedure described above, and represented in the circuit in Fig.~\ref{fig:dict_with_dagger}, we can
    compute the inner product of the states in Equations~\ref{eqn:state_after_f} and~\ref{eqn:h_and_unitary_b}:

    \begin{equation*}
        \begin{split}
            E_{\ket{0}} & \coloneqq \bra{0}_{n+m} (H^{\otimes n} \otimes B^{\dagger}) F (A \otimes I_m)
            \ket{0}_{n+m} \\
            & = \frac{1}{\sqrt{N}} \sum_{k=0}^{N-1}a_k b_{f(k)} \\
            & = \frac{ab}{\sqrt{N}} \sum_{k=0}^{N-1}w_k h_{f(k)}.
        \end{split}
    \end{equation*}

    Therefore,

    \begin{equation*}
            \sum_{k=0}^{N-1}w_k h_{f(k)} = \frac{\sqrt{N}}{ab} E_{\ket{0}}
    \end{equation*}

    with $E_{\ket{0}} = \bra{0}_{n+m} (H^{\otimes n} \otimes B^{\dagger}) F (A \otimes I_m)\ket{0}_{n+m}$ being the
    amplitude of $\ket{0}_{n+m}$ at the end of the computation.

    The desired weighted sum is the amplitude of the initial state $\ket{0}_{n+m}$ after applying the unitary
    operator $(H^{\otimes n} \otimes B^{\dagger}) F (A \otimes I_m)$ to it, which can be estimated using
    amplitude estimation algorithms.
\end{proof}

\begin{application1}
    If $B$ is the operator $L_m$ that encodes $h$ as the identity, i.e. $h(v) = v$ for $0 \le v < M$, and $b =
    \frac{1}{\sqrt{\sum_{v=0}^{M-1} v^2}} = \sqrt{\frac{6}{(M-1)M(2M-1)}}$ we obtain a canonical way to compute the
    expected value of $f$:

    \begin{equation*}
        \sum_{k=0}^{N-1}w_k f(k) = \frac{\sqrt{N}}{a} \sqrt{\frac{(M-1)M(2M-1)}{6}} E_{\ket{0}}
    \end{equation*}

    with $E_{\ket{0}} = \bra{0} (H^{\otimes n} \otimes L_m^{\dagger}) F(A\otimes I_m)\ket{0}_{n+m}$ being the
    amplitude of $\ket{0}_{n+m}$ at the end of the computation.
\end{application1}

\begin{application3}
    With the notations in the \emph{\textbf{Expected Value}} context, if $A = H^{\otimes n}$, then $w_k = 1$ for
    $0 \le k < N$ and $a = \frac{1}{\sqrt{N}}$, and we get the mean value of $f$:

    \begin{equation*}
        \frac{1}{N} \sum_{k=0}^{N-1} f(k) = \sqrt{\frac{(M-1)M(2M-1)}{6}} E_{\ket{0}}
    \end{equation*}

    with $E_{\ket{0}} = \bra{0} (H^{\otimes n} \otimes L_m^{\dagger}) F(H^{\otimes n} \otimes I_m)\ket{0}_{n+m}$
    being the amplitude of $\ket{0}_{n+m}$ at the end of the computation.
\end{application3}

\begin{application2}
    For subsets $K \subseteq \{0, \mathellipsis, N-1\}$ and $V \subseteq \{0, \mathellipsis, M-1\}$, we can compute
    the restricted weighted sum,

    \begin{equation*}
        \sum_{\substack{k \in K \\ f(k) \in V}} w_k f(k),
    \end{equation*}

    by appropriate choices of $f$ and $h$. In particular, value at risk (VaR) and
    comparators needed in option pricing can be computed using this pattern.
\end{application2}

Note that Lebesgue integrals and partial moments of statistical
distributions are also applications of the generalized inner product.

%%%%%%%%%%%%%%%%%%%%%%%%%%%%%%%%%%%%%%%%%%%%%%%%%%%%%%%%%%%%%%%%%%%%%%%%%%%
\section{\label{sec:distributions}Quantum State Encoding of Common Distributions and Functions}
%%%%%%%%%%%%%%%%%%%%%%%%%%%%%%%%%%%%%%%%%%%%%%%%%%%%%%%%%%%%%%%%%%%%%%%%%%%

Efficient quantum state preparation is critical for computing inner products. Specifically, encoding a normal
distribution as a quantum state is computationally notoriously difficult ~\cite{HerbertNormal}. The same is true for
the preparation of quantum states with amplitudes or probabilities corresponding to the values of a linear function
or higher order polynomial. For example, in~\cite{QuantumRisk} a value $y$ of a polynomial is approximated by $y
\approx \sin^2(y + \frac{\pi}{4}) - \frac{1}{2}$, which is equivalent to $y \approx \frac{1}{2}\sin(2y)$ for $y \approx 0$.
Note that this is an approximation using probabilities. More details about this approximation are provided in the
Appendix~\ref{sec:linear_sw}.

In this section we introduce an exact encoding of the raised cosine distribution, in probabilities,
(Section~\ref{subsec:raised_cosine}), which can approximate the normal distribution ~\cite{GreenNormalApprox}. The
implementation borrows an idea from Digital Signal Processing. We encode Fourier coefficients in the quantum state,
and then apply the (inverse) quantum Fourier transform. We use this same idea to provide increasingly better
approximations for the normal distribution for a given number of qubits using more Fourier coefficients in
Section~\ref{subsec:normal_sin4}.

Additionally, in Section~\ref{subsec:linear_approx_state_prep} we use a canonical encoding of a linear function using
the approximation $y\approx\sin(y)$, for $y \approx 0$, in amplitudes instead of probabilities. We also provide a
heuristic implementation for a high-precision approximation of identity as a canonical implementation of a linear
function for three, four and five qubits (Section~\ref{subsec:identity}).

\subsection{\label{subsec:raised_cosine}Exact Encoding of the Raised Cosine Distribution}

In this section, we describe a quantum implementation of the raised cosine probability distribution. The probability
density function for the raised cosine distribution is

\begin{equation}
    \label{eqn:raised_cosine_density}
    \begin{split}
        p(x) & = \frac{1}{2\sigma}\left(1 + \cos\left(\frac{x-\mu}{\sigma}\pi\right) \right) \\
        & = \frac{1}{\sigma}\cos^2\left (\frac{x-\mu}{2\sigma}\pi\right) \\
    \end{split}
\end{equation}

for $\mu - \sigma \le x \le \mu + \sigma$.\\
\bigskip

\begin{figure}[ht]
    \centering
    \mbox{
        \Qcircuit @C=1em @R=0.0em @!R {
            & \lstick{\ket{q}_0{}}          & \qw     & \qw         & \qw            &\multigate{3}{QFT^{\dagger}}  &\qw  \\
            & \lstick{\ket{q}_1{}}          & \qw     & \qw         & \qw            &\ghost{QFT^{\dagger}}         & \qw  \\
            & \lstick{\textcolor{white}{0}} &\vdots                                                                                 \\
            & \lstick{\ket{q}_{n-1}{}}      & \qw     & \gate{H}    & \gate{P(\pi)}  &\ghost{QFT^{\dagger}}         & \qw  \\
        }
    }
    \caption{The quantum circuit that prepares the state with a raised cosine probablity distribution represented in Eq
    .~\ref{eqn:raised_cosine_state}, where $n$ is a positive integer representing the number of qubits in the
    system, $H$ is a single qubit Hadamard gate, $P$ is a single qubit Phase gate.}
    \label{fig:raised_cosine_circuit}
\end{figure}
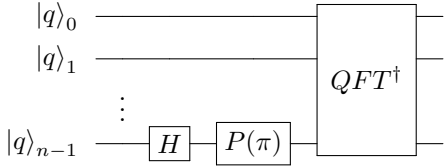

Given a quantum system with $n$ qubits, we start by encoding the state

\begin{equation*}
    \frac{1}{\sqrt{2}} \ket{0}_n - \frac{1}{\sqrt{2}} \ket{2^{n-1}}_n
\end{equation*}

then we apply the inverse quantum Fourier transform. The final circuit, denoted by $N_{1, n}$, is represented in
Fig.~\ref{fig:raised_cosine_circuit}. The resulting state is

\begin{equation}
    \label{eqn:raised_cosine_state}
    \begin{split}
        \ket{\nu_1}_n & = N_{1, n} \ket{0}_n \\
        & = \sqrt{\frac{2}{N}}\sum_{k = 0}^{N-1}\sin(k\frac{\pi}{N})e^{i(\frac{\pi}{2} - k\frac{\pi}{N})}\ket{k}_n.
    \end{split}
\end{equation}

The corresponding probability distribution is

\begin{equation}
    \label{eqn:raised_cosine_probs}
    \begin{split}
        p(k) & = \frac{2}{N}\cos^2\left(\left(k-\frac{N}{2}\right)\frac{\pi}{N}\right) \\
        & = \frac{2}{N} \sin^2\left (k\frac{\pi}{N}\right)
    \end{split}
\end{equation}

for $0 \le k < N$. This matches the probability density function of the raised cosine distribution
(Equation~\ref{eqn:raised_cosine_density}) when $\mu = \frac{N}{2}$ and $\sigma = \frac{N}{2}$.

\begin{figure}
    \centering
    \includegraphics[width=.45\textwidth]{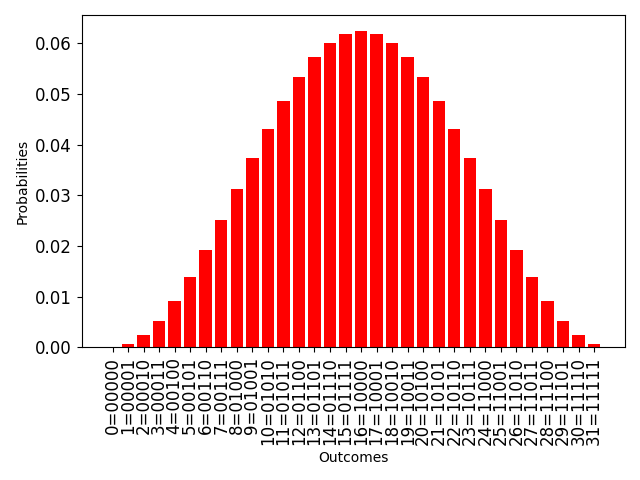}
    \caption{Outcome probability distribution after encoding the raised cosine distribution using a five-qubit
    quantum state.}
    \label{fig:raised_cosine_probs}
\end{figure}

\subsection{\label{subsec:normal_sin4}Approximations of the Normal Distribution with Trigonometric Functions}

The raised cosine distribution is an approximation for the normal distribution, as introduced by Raab  and Green
in~\cite{GreenNormalApprox}.

We can use the equivalent of three Fourier coefficients instead of two and create the
$n$-qubit quantum state:

\begin{equation}
    \label{eqn:fourier_coeffs_3}
    \sqrt{\frac{2}{3}} \ket{0}_n - \frac{1}{\sqrt{6}} \ket{2^{n-1}}_n - \frac{1}{\sqrt{6}} \ket{2^n - 1}_n
\end{equation}

Then, we apply the inverse quantum Fourier transform. Let us denote the composite unitary operator by $N_{2, n}$:

The resulting state is

\begin{equation}
    \label{eqn:sin4_state}
    \begin{split}
        \ket{\nu_2}_n & = N_{2, n} \ket{0}_n\\
        & = \sqrt{\frac{8}{3N}}\sum_{k = 0}^{N-1}\sin^2\left(k\frac{\pi}{N}\right) \ket{k}_n.
    \end{split}
\end{equation}

The corresponding probability distribution is

\begin{equation}
    \label{eqn:sin4_distribution}
        p(k) = \frac{8}{3N}\cos^4\left(\left(k-\frac{N}{2}\right)\frac{\pi}{N}\right) = \frac{8}{3N} \sin^4\left
        (k\frac{\pi}{N}\right)
\end{equation}

for $0 \le k < N$.

Note that:

\begin{equation}
    \label{eqn:sum_sin2}
    \sum_{k=0}^{N-1}\sin^2\left(k\frac{\pi}{N}\right) = \frac{N}{2}
\end{equation}

The resulting probability distribution is a better approximation of a normal distribution than the raised
cosine probability distribution. Using more Fourier coefficients yields even better approximations.

We can use five Fourier coefficients to create the state

\begin{equation}
    \label{eqn:fourier_coeffs_5}
    \begin{split}
        \frac{6}{\sqrt{70}} & \ket{0}_n + \frac{1}{\sqrt{70}} \ket{2^{n-2}}_n + \frac{1}{\sqrt{70}} \ket{2^{n-1} -1}_n\\
        & - \frac{4}{\sqrt{70}} \ket{2^{n - 1}}_n - \frac{4}{\sqrt{70}} \ket{2^n - 1}_n,
    \end{split}
\end{equation}

followed by the inverse quantum Fourier transform. Let us denote the composite unitary operator by $N_{4, n}$:

\begin{equation*}
    \label{eqn:sin8_state}
    \begin{split}
        \ket{\nu_4}_n & = N_{4, n} \ket{0}_n \\
        & = \sqrt{\frac{128}{35N}}\sum_{k = 0}^{N-1}\sin^4\left (k\frac{\pi}{N}\right) \ket{k}_n
    \end{split}
\end{equation*}

The corresponding probability distribution is

\begin{equation}
    \label{eqn:sin8_distribution}
    p(k) = \frac{128}{35N}\cos^8\left(\left(k-\frac{N}{2}\right)\frac{\pi}{N}\right) = \frac{128}{35N}
    \sin^8\left (k\frac{\pi}{N}\right)
\end{equation}

for $0 \le k < N$.

\begin{figure*}[ht]
    \centering
    \begin{minipage}{.45\textwidth}
        \centering
        \includegraphics[width=.9\linewidth]{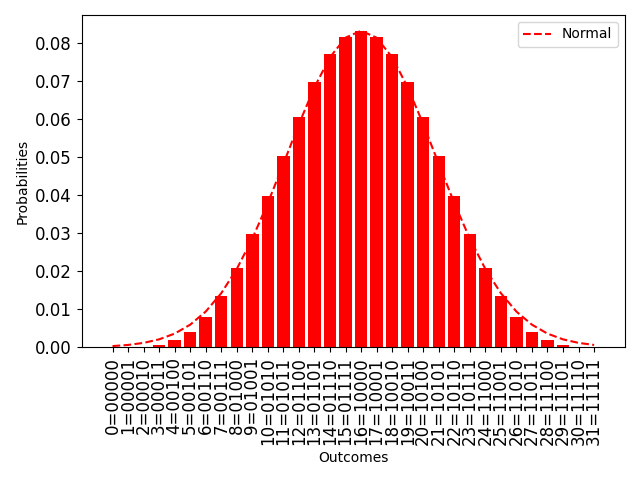}
    \end{minipage}
    \begin{minipage}{.45\textwidth}
        \centering
        \includegraphics[width=.9\linewidth]{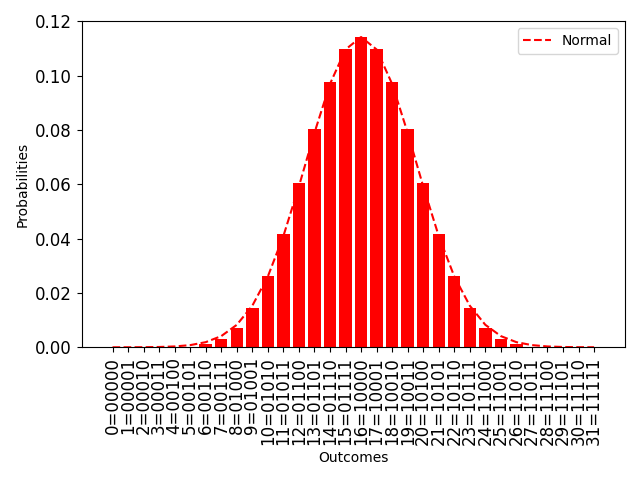}
    \end{minipage}
    \caption{Left: The outcome probability distribution of the five-qubit state prepared using three Fourier
    coefficients (Eq.~\ref{eqn:fourier_coeffs_3}) resulting in the probability distribution in Eq
    .~\ref{eqn:sin4_distribution}. The corresponding normal distribution is included for comparison. Right: The
    outcome probability distribution of the five-qubit state prepared using five Fourier coefficients (Eq
    .~\ref{eqn:fourier_coeffs_5}) resulting in the probability distribution in Eq.~\ref{eqn:sin8_distribution}. The
    corresponding normal distribution is included for comparison.}
    \label{fig: sin4_amps_probs}
\end{figure*}

\subsection{\label{subsec:linear_approx_state_prep}Approximation of Linear Functions with Trigonometric Functions}

Given a quantum system with $n$ qubits and any $\theta \in \mathbb{R}$, let us denote by $T_{\theta, n}$ a unitary
operator that creates the state

\begin{equation}
    \label{eqn:linear_approx_trig}
    \begin{split}
        & \ket{\tau_{\theta}}_{n+1} = T_{\theta, n} \ket{0}_n \\
        & = \frac{1}{\sqrt{N}}\sum_{k = 0}^{N-1}\sin(k\theta) \ket{k}_n\ket{0}_1 + \frac{1}{\sqrt{N}}\sum_{k =
        0}^{N-1}\cos(k\theta)\ket{k}_n\ket{1}_1
    \end{split}
\end{equation}

For a small $\theta$ the amplitudes $\frac{1}{\sqrt{N}} \sin(k\theta)$ approximate the values of the linear function $k
\mapsto \frac{\theta}{\sqrt{N}}k$ for $0 \le k < N$.

\begin{figure}[ht]
    \centering
    \mbox{
        \Qcircuit @C=1em @R=0.0em @!R {
            & \lstick{\ket{q}_0{}}          & \qw     & \gate{H}    & \ctrl{3}     &\qw \\
            & \lstick{\ket{q}_1{}}          & \qw     & \gate{H}    & \ctrl{2}     &\qw  \\
            & \lstick{\textcolor{white}{0}} &\vdots \\
            & \lstick{\ket{q}_{n-1}{}}      & \qw     & \gate{X}    & \gate{Ry(\theta)} &\qw\\
        }
    }
    \label{fig:linear_approx_circuit}
    \caption{The quantum circuit that prepares the quantum state represented in Eq.~\ref{eqn:linear_approx_trig}.}
\end{figure}
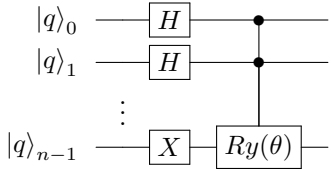

\subsection{\label{subsec:identity}Heuristic Implementation of Linear Functions}

Given a number of qubits, non-canonical, heuristic search methods can be used to create arbitrarily close
approximations of linear functions.

For an $n$ qubit quantum system, with $N = 2^n$, let us denote by $L_n$ a unitary operator that creates the state

\begin{equation}
    \label{eqn:heuristic_lin}
    \begin{split}
        \ket{\lambda}_n & = L_n \ket{0}_n \\
        & = \sqrt{\frac{6}{(N-1)N(2N-1)}}\sum_{k = 0}^{N-1}k\ket{k}_n.
    \end{split}
\end{equation}

Visualizations of the quantum states $\ket{\lambda}_3$, $\ket{\lambda}_4$ and $\ket{\lambda}_5$ are in
Appendix~\ref{subsec:catalog_linear_heuristic}.

%%%%%%%%%%%%%%%%%%%%%%%%%%%%%%%%%%%%%%%%%%%%%%%%%%%%%%%%%%%%%%%%%%%%%%%%%%%
\section{\label{sec:applications}Applications}
%%%%%%%%%%%%%%%%%%%%%%%%%%%%%%%%%%%%%%%%%%%%%%%%%%%%%%%%%%%%%%%%%%%%%%%%%%%

In this section we are presenting applications of the Weighted Sum pattern~\ref{pattern1} and
Controlled Weighted Sum pattern~\ref{pattern2} to expected value, value at risk, payoff, and value counting
computations used in financial engineering and optimization.

Functions need to be discretized when encoded in quantum states. The Quantum Dictionary pattern allows to
encode discrete functions represented as binary polynomials as described in Section~\ref{subsec:binary_poly_encoding}.
In~\cite{StamatopoulosOptionPricing} it is shown that the expected value of any linear function can be computed
using a canonical linear function. Appendix~\ref{woerner_egger} contains a short review of this
approximation method.

We show that having an efficient implementation of a canonical linear function
(e.g. the identity) allows the computation of the expected value of any discrete function. The implementation can
be exact or approximate.

\subsection{\label{subsec:expected_value_poly}The Expected Value of Any Discrete Function}

In this section we assume the availability of an operator $B$ that prepares the state defined in
Equation~\ref{eqn:heuristic_lin}.

Given integers $n > 0$ and $m > 0$, with $N = 2^n$ and $M = 2^m$, and a function $f: \{0, \mathellipsis, 2^n-1\}
\rightarrow \{0, \mathellipsis, 2^m-1\}$, assume we want to compute the weighted sum

\begin{equation*}
    \sum_{k=0}^{N-1}w_k f(k)
\end{equation*}

for weights $w_k = \sin^2(k\frac{\pi}{N})$ for $0 \le k < N$.

Considering the representation of $f$ as a binary polynomial $p:\{0, 1\}^n \rightarrow \{0, 1\}^m$, as described in
Section~\ref{subsec:function_poly}, we need to compute the weighted sum

\begin{equation*}
    \sum_{k=0}^{N-1}w_k p(k).
\end{equation*}

As an example, with $n = 3$ and $m = 4$, we will show how to perform this computation for the binary polynomial

\begin{equation}
    \label{eqn:bin_poly_expected_value}
    p(k_0, k_1, k_2) = 7 + 4 k_1 - 5 k_0 k_1 - 2 k_0 k_2
\end{equation}

for $(k_0, k_1, k_2) \in \{0, 1\}^n$, with $k = \sum_{j=0}^{n-1}k_j 2^j$ being the binary expansion of $k$ as in
Section~\ref{subsec:function_poly}.

Using the Controlled Weighted Sum pattern~\ref{pattern2}, where $A = N_{2,n}$ as described in
Equation~\ref{eqn:sin4_state}, and $B = L_m$ as described in Equation~\ref{eqn:heuristic_lin}, $a =
\sqrt{\frac{8}{3N}}$, and $b = \sqrt{\frac{6}{(M-1)M(2M-1)}}$ we obtain the result

\begin{equation}
    \label{eqn:expected_discrete}
    \begin{split}
        \sum_{k=0}^{N-1}w_k f(k) & = \sqrt{N} \sqrt{\frac{3N}{8}} \sqrt{\frac{(M-1)M(2M-1)}{6}} \\
        & \cdot \bra{0} (H^{\otimes n} \otimes B^{\dagger}) F (A \otimes I_m)\ket{0}_{n+m}
    \end{split}
\end{equation}

\begin{figure}[ht]
    \begin{tabular}{c}
        \includegraphics[width=0.95\linewidth]{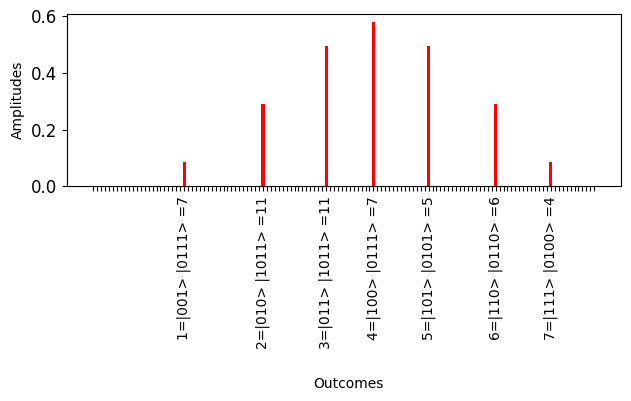} \\
        \includegraphics[width=0.95\linewidth]{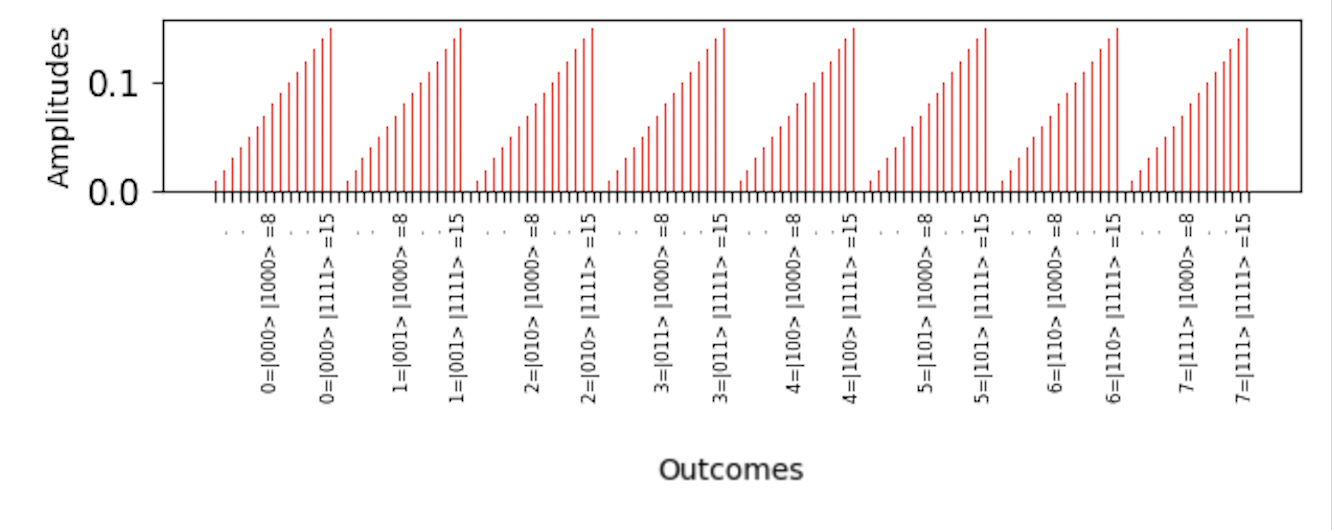} \\
    \end{tabular}
    \captionof{figure}{Top: Visualization of the amplitudes of a quantum system after applying the operator $A =
    N_{2,3}$, using 3 qubits for the key register and and 4 qubits for the value register. The key and  value pairs
    on the x-axis show the result of encoding the binary polynomial in Eq.~\ref{eqn:bin_poly_expected_value}. The
    distribution of weights applied to the key register (as in Eq.~\ref{eqn:sin4_state}) is an approximation of a
    normal distribution in the amplitudes. Bottom: Visualization of amplitudes of a quantum system, using 3 qubits
    for the key register and 4 qubits for the value register, after applying the operator $B = L_4$ and Hadamard
    gates on the key register. Encoding the linear function in the value register (as in Eq
    .~\ref{eqn:heuristic_lin}) results in the identity function implemented in amplitudes, repeated for each key
    value.}
    \label{fig:dict_normal_identity}
\end{figure}

Running the quantum computation in a simulator yields $\bra{0} (H^{\otimes n} \otimes B^{\dagger}) F (A \otimes I_m)
\ket{0}_{n+m} = 0.17835$ (the amplitude of $\ket{0}_{n+m}$), and from Equation~\ref{eqn:expected_discrete} we obtain
$\sum_{k=0}^{N-1}w_kf(k) \approx 30.76777$.  A direct classical calculation gives $\sum_{k=0}^{N-1}w_kf(k) \approx
30.76777$.

The results of performing this experiment on real quantum hardware can be found in Section~\ref{sec:experiments}.

\subsection{\label{subsec:payoff}Payoff Computation in Option Pricing}

For simplicity we will use the same discrete function in Section~\ref{subsec:expected_value_poly} to represent
a payoff function in an option pricing calculation, whose values are listed below.

\begin{center}
    \begin{tabular}{c|c}
        $k$ & $f(k)$ \\
        \hline
        $0$ & $7$    \\
        $1$ & $7$    \\
        $2$ & $11$   \\
        $3$ & $11$   \\
        $4$ & $7$    \\
        $5$ & $5$    \\
        $6$ & $6$    \\
        $7$ & $4$    \\
    \end{tabular}
\end{center}

Assume that the strike price is $K=7$, and the price distribution is represented by the values $w_k$. Then
we are interested in computing

\begin{equation*}
    \sum_{\substack{0 \le k < N \\ f(k) \ge K}} w_k \left(f(k) - K\right)
\end{equation*}

We can do that by using the Controlled Weighted Sum pattern~\ref{pattern2} with hashes

\begin{equation*}
    h_v =
    \begin{cases}
        v-K, & \text{if } v \ge K\\
        0, & \text{otherwise}\\
    \end{cases}
\end{equation*}

Alternatively, we can encode the function $f - K$ instead of $f$ and choose hashes that are identity for non-negative
inputs and $0$ for negative ones. Note that the hash function acts as a
value selector and replaces the comparator used in~\cite{StamatopoulosOptionPricing}.

\subsection{\label{subsec:var}Value at Risk}

Within the context and with the notations of the Weighted Sum pattern~\ref{pattern1}, for a given $0 \le l < N$,
if $b = \frac{1}{\sqrt{l}}$ and

\begin{equation*}
    f(k) =
    \begin{cases}
        1, & \text{if } k \le l\\
        0, & \text{otherwise}\\
    \end{cases}
\end{equation*}

then we can compute

\begin{equation*}
    \label{eqn:quantile}
    \sum_{k=0}^l w_k.
\end{equation*}

This allows the computation of Value at Risk when the values $w_k$ represent a price distribution
by using a binary search over $l$ as also described in~\cite{QuantumRisk}.

Fig.~\ref{fig:quantile} shows an example of a state that can be used to estimate the sum in
Equation~\ref{eqn:quantile} for $N=8$ and $l=3$

\begin{center}
    \includegraphics[width=0.95\linewidth]{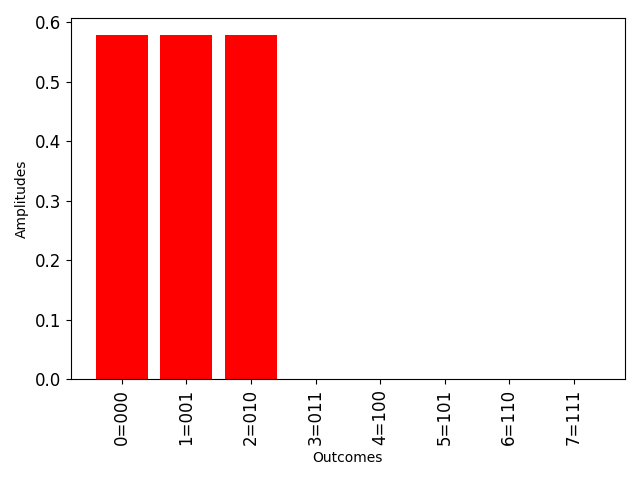}
    \captionof{figure}{Visualization of the amplitudes of a three-qubit quantile quantum state, used in computing
    Value at Risk for $N=8$ and $l=3$.}
    \label{fig:quantile}
\end{center}

Note that the Controlled Weighted Sum pattern~\ref{pattern2} can also be used to estimate Value at Risk.

\subsection{\label{subsec:expected_value_rational}The Expected Value of a Non-Linear Function}

Next, we consider the computation of the expected value of a non-linear function with a normally distributed argument.

Given an integer $n > 0$ and $N = 2^n$, assume we want to compute the weighted sum of a function $f$ defined by
$f(k) = r(x, y)$, for integers $0 \le k < N$,  where $k = 4x + y$, $0 \le x,y < 3$:

\begin{equation*}
    \sum_{k=0}^{N-1}w_k f(k)
\end{equation*}

for weights $w_k = \sin^2(k\frac{\pi}{N})$ for $0 \le k < N$, where $r$ is the rational function defined by

\begin{equation*}
    r(x, y) =  \frac{1}{7.856} \left (\frac{4.01-x}{1+x} + \frac{4.01-2y+x}{(1+y)^{2}} - 0.344 \right )
\end{equation*}

The rational function $r$ is an example of a function that can be used in derivative pricing.

Using the Weighted Sum pattern ~\ref{pattern1}, where $A = N_{2,4}$, as described in Equation~\ref{eqn:sin4_state},
for $n = 4$ qubits, and $B$ encodes normalized values of the function $f$, $a = \sqrt{\frac{8}{3N}}$ and $b = 1$,
we obtain the result

\begin{equation*}
    \sum_{k=0}^{N-1}w_k f(k) = \sqrt{\frac{3N}{8}} \bra{0}B^{\dagger}A \ket{0}_{n}
\end{equation*}

The states prepared by operators $A$ and $B$ are shown in Fig.~\ref{fig:a_and_b_rational}.

\begin{center}
    \begin{tabular}{c}
        \includegraphics[width=0.95\linewidth]{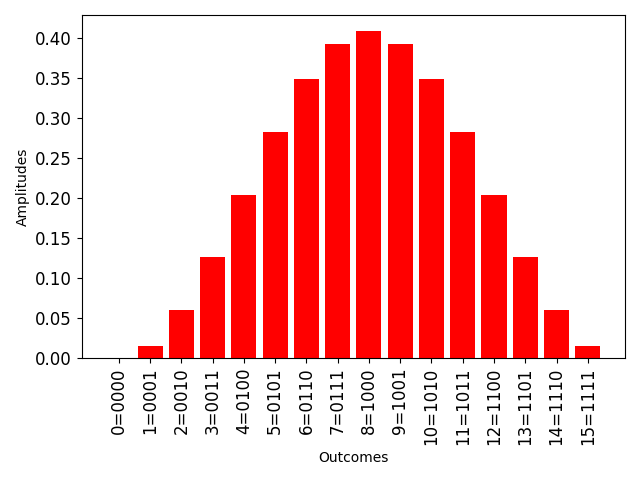} \\
        \includegraphics[width=0.95\linewidth]{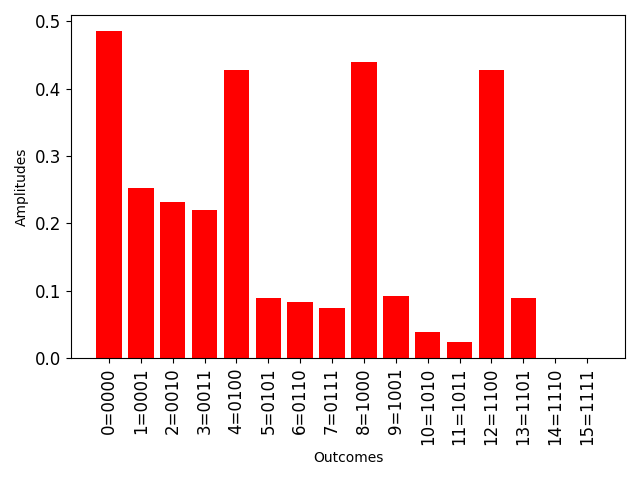} \\
    \end{tabular}
    \captionof{figure}{Top: Visualization of the amplitudes of a four-qubit quantum state prepared by the
    unitary operator $A$, as described in Eq.~\ref{eqn:sin4_state}. Bottom: Visualization of the amplitudes of a
    four-qubit quantum state prepared by the unitary operator $B$, which encodes normalized values of the
    function $f$ as described in Section~\ref{subsec:expected_value_rational}.}
    \label{fig:a_and_b_rational}
\end{center}

Running the computation in a simulator yields $\bra{0}B^{\dagger}A \ket{0}_{n} \approx 0.55050$, leading to
$\sum_{k=0}^{N-1}w_k f(k) \approx 1.34845$. A direct classical calculation gives $\sum_{k=0}^{N-1}w_k f(k) \approx
1.33431$.

\subsection{\label{subsec:expected_value_linear}The Expected Value of Linear Functions: Exact Version}

\subsubsection{\label{subsubsec:linear_exact_1}Any Linear Function}

Next, we consider the computation of the expected value of the identity function, with a normally distributed argument.

Given an integer $n > 0$ and $N = 2^n$, assume we want to compute the weighted sum of a function $f$ defined by
$f(k) = 1 + 2k$ for integers $0 \le k < N$:

\begin{equation*}
    \sum_{k=0}^{N-1}w_kf(k)
\end{equation*}

for weights $w_k = \sin^2(k\frac{\pi}{N})$ for $0 \le k < N$.

Using the Weighted Sum pattern~\ref{pattern1}, where $A = N_{2, 3}$ as described in  Equation~\ref{eqn:sin4_state},
for $n = 3$ qubits, $B = L_3$ as described in Equation~\ref{eqn:heuristic_lin}, $a = \sqrt{\frac{8}{3N}}$ and $b =
\frac{1}{\sqrt{\sum_{k = 0}^{N-1}(1 + 2k)^2}}$, we obtain

\begin{equation*}
    \sum_{k=0}^{N-1}w_k k = \sqrt{\frac{3N}{8}} \sqrt{\sum_{k = 0}^{N-1}(1 + 2k)^2} \bra{0}B^{\dagger}A \ket{0}_{n}.
\end{equation*}

The states prepared by operators $A$ and $B$ are shown in Fig.~\ref{fig:a_and_b_1+2k}.

\begin{center}
    \begin{tabular}{c}
        \includegraphics[width=0.8\linewidth]{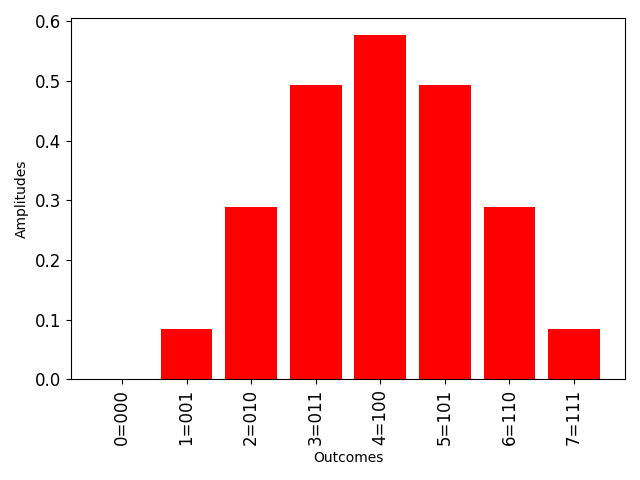} \\
        \includegraphics[width=0.8\linewidth]{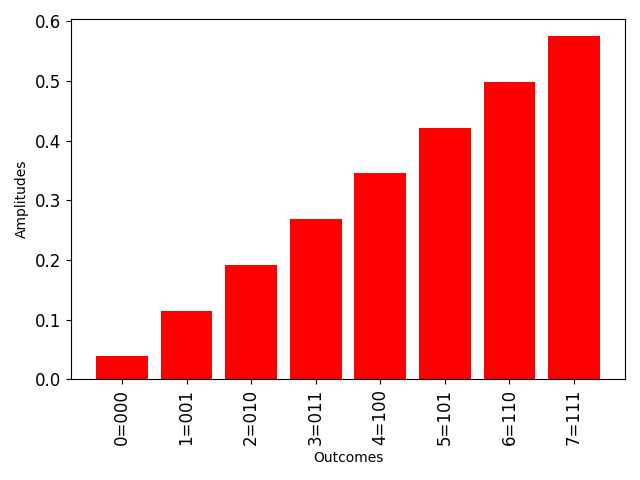} \\
    \end{tabular}
    \captionof{figure}{Top: Visualization of the amplitudes of a quantum state prepared by the unitary operator $A =
    N_{2,3}$, as described in Eq.~\ref{eqn:sin4_state}, using three qubits. Bottom: Visualization of the amplitudes
    of a quantum state prepared by the unitary operator $B = L_3$, as described in Eq.~\ref{eqn:heuristic_lin}, using
    three qubits.}
    \label{fig:a_and_b_1+2k}
\end{center}

As an example, for $n = 3$ and $N = 2^n$, running the computation in a simulator gives $\bra{0}B^{\dagger}A
\ket{0}_{n} \approx 0.79705$, leading to $\sum_{k=0}^{N-1}w_k (1 + 2k) \approx
36.0$. A direct computation shows that $\sum_{k=0}^{N-1}w_k f(k) = 36.0$.

\subsubsection{\label{subsubsec:linear_exact_2}A Canonical Linear Function}

Given an integer $n > 0$ and $N = 2^n$, assume we want to compute the weighted sum of the identity function $f$
defined by $f(k) = k$ for integers $0 \le k < N$:

\begin{equation*}
    \sum_{k=0}^{N-1}w_k k
\end{equation*}

for weights $w_k = \sin^2(k\frac{\pi}{N})$ for $0 \le k < N$.

Using the Weighted Sum pattern~\ref{pattern1}, where $A = N_{2, 3}$ as described in  Equation~\ref{eqn:sin4_state},
for $n = 3$ qubits, $B = L_3$ as described in Equation~\ref{eqn:heuristic_lin}, $a = \sqrt{\frac{8}{3N}}$, and $b =
\sqrt{\frac{6}{(N-1)N(2N-1)}}$, we obtain

\begin{equation*}
    \sum_{k=0}^{N-1}w_k k = \frac{N}{4} \sqrt{(N-1)(2N-1)} \bra{0}B^{\dagger}A \ket{0}_{n}
\end{equation*}

The states prepared by operators $A$ and $B$ are shown in Fig.~\ref{fig:a_and_b_3}.

\begin{center}
    \begin{tabular}{c}
        \includegraphics[width=0.8\linewidth]{sin4_q3.png} \\
        \includegraphics[width=0.8\linewidth]{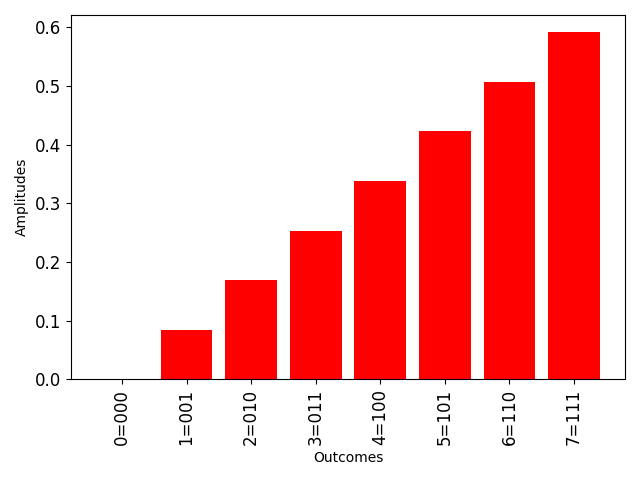} \\
    \end{tabular}
    \captionof{figure}{Top: Visualization of the amplitudes of a three-qubit quantum state prepared by the
    unitary operator $A$, as described in Eq.~\ref{eqn:sin4_state}. Bottom: Visualization of the amplitudes of a
    three-qubit quantum state prepared by the unitary operator $B$, as described in Eq.~\ref{eqn:heuristic_lin}.}
    \label{fig:a_and_b_3}
\end{center}

Running the computation in a simulator gives $\bra{0}B^{\dagger}A \ket{0}_{n} \approx 0.77998$,
leading to $\sum_{k=0}^{N-1}w_k k \approx 15.98493$.

Note that this canonical implementation of a linear functions allows for the computation of the expected value of any
other linear function. For example, we can compute the weighted sum of the values of the function $f$ defined by $f
(k) = 1 + 2k$ for integers $0 \le k < N$:

\begin{equation*}
    \sum_{k=0}^{N-1} w_k f(k)
\end{equation*}

for weights $w_k = \sin^2(k\frac{\pi}{N})$ for $0 \le k < N$.

\begin{equation*}
    \begin{split}
        & \sum_{k=0}^{N-1}w_k f(k) = \sum_{k=0}^{N-1}w_k (1+2k) \\
        & = \sum_{k=0}^{N-1}w_k + 2\sum_{k=0}^{N-1}w_k k = \frac{N}{2} + 2\sum_{k=0}^{N-1}w_k
    \end{split}
\end{equation*}

For $n=3$, and therefore $N=8$, we obtain  $\sum_{k=0}^{N-1}w_k (1 + 2k) \approx 4 + 2 \cdot 15.98493 = 35.96986$.

\subsection{\label{subsec:expected_linear_approx}The Expected Value of Linear Functions: Approximate Version}

This section is inspired by the method used in~\cite{StamatopoulosOptionPricing}, except the approximation is done
using amplitudes instead of probabilities.

As in Section~\ref{subsec:expected_value_linear}, for a given an integer $n > 0$ and $N = 2^n$,
assume we want to compute the weighted sum of the identity function $f$
defined by $f(k) = k$ for integers $0 \le k < N$:

\begin{equation*}
    \sum_{k=0}^{N-1}w_k k
\end{equation*}

for weights $w_k = \sin^2(k\frac{\pi}{N})$ for $0 \le k < N$.

We will rely on the canonical approximation from Section
~\ref{subsec:linear_approx_state_prep}.

Using the Weighted Sum pattern~\ref{pattern1}, where $A = N_{2,4}$, as described in Equation~\ref{eqn:sin4_state},
and $B = T_{\frac{c}{2N}, 4}$, as described in Equation~\ref{eqn:linear_approx_trig}, for $n = 4$ and a value $c \approx 0$.
Then $a = \sqrt{\frac{8}{3N}}$, $b = \sqrt{\frac{1}{N}}\frac{c}{2N}$, and we obtain

\begin{equation*}
    \begin{split}
        \sum_{k=0}^{N-1}w_k k & = \frac{1}{\sqrt{\frac{8}{3N}}\sqrt\frac{1}{N}\frac{c}{2N}} \bra{0}B^{\dagger}A
        \ket{0}_{n+1} \\
        & = \sqrt{\frac{3}{2}}\frac{N^2}{c}\bra{0}B^{\dagger}A \ket{0}_{n+1}
    \end{split}
\end{equation*}

The states prepared by operators $A$ and $B$ are shown in Fig.~\ref{fig:a_and_b_approx}.
The circuit will have $n$-qubits to encode the input $0 \le k < N$ and an ancillary qubit needed as target for
the $R_Y$ rotations in the operator $B$.

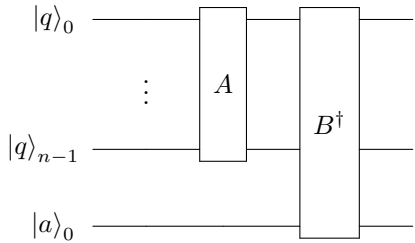
\begin{figure}[ht]
    \centering
    \mbox{
        \Qcircuit @C=2em @R=2em {
            & \lstick{\ket{q}_0{}}      	& \qw   & \multigate{2}{A} 	& \multigate{3}{B^{\dagger}}  	& \qw    \\
            & \lstick{\textcolor{white}{0}} & \vdots \\
            & \lstick{\ket{q}_{n-1}}      	& \qw   & \ghost{A}      	& \ghost{B^{\dagger}}         	& \qw    \\
            & \lstick{\ket{a}_0{}}      	& \qw   & \qw               & \ghost{B^{\dagger}} 			& \qw    \\
        }
    }
    \caption{\label{fig:dagger_ancilla}The quantum circuit that computes the inner product represented in
    Eq.~\ref{eqn:inner_prod_def} using unitary operators $A$ and $B^\dagger$ (as defined in
    Eq.~\ref{eqn:transformations_a_and_b}), and an ancillary qubit as the target for the $R_y$ rotations in
    operator $B^\dagger$.}
\end{figure}

\begin{center}
    \begin{tabular}{c}
        \includegraphics[width=0.95\linewidth]{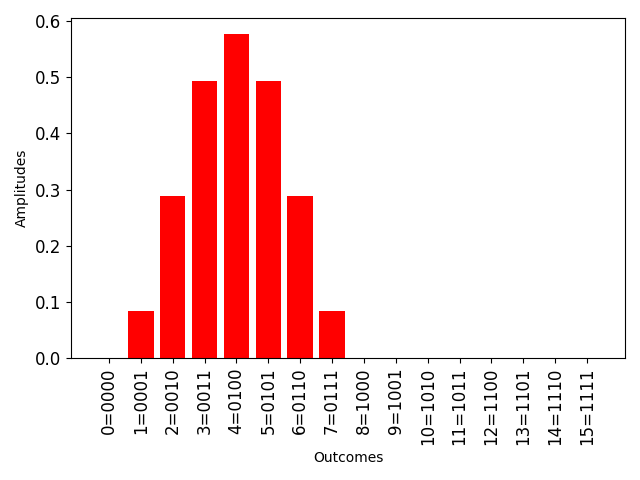} \\
        \includegraphics[width=0.95\linewidth]{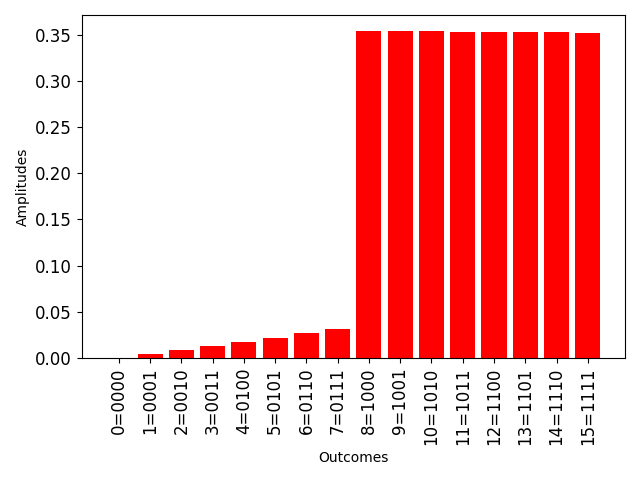} \\
    \end{tabular}
    \captionof{figure}{Top: Visualization of the amplitudes of a three-qubit quantum state prepared by the unitary
    operator $A$, as described in Eq.~\ref{eqn:sin4_state}. Bottom: Visualization of the amplitudes of a
    three-qubit quantum state prepared by the unitary operator $B$, as described in Eq
    .~\ref{eqn:linear_approx_trig}, with a ancillary qubit added to the circuit to serve as the target
    for $R_y$ rotations.}
    \label{fig:a_and_b_approx}
\end{center}

As an example, for $n = 3$, $N = 2^n$ and $c = 0.1$, when running the quantum circuit in a simulator we measure
the amplitude of $\ket{0}_{n+1}$ as
$\bra{0}B^{\dagger}A \ket{0}_{n+1} \approx 0.02041$ (it depends only on
the distribution), leading to the approximation:

\begin{equation*}
    \sum_{k=0}^{N-1} w_k k = \sqrt{\frac{3}{2}}\frac{N^2}{c} 0.02041 = 15.99768.
\end{equation*}

Like in Section~\ref{subsubsec:linear_exact_2}, we can use this canonical implementation
to compute the expected value of any
other linear function, e.g. $f$ defined by $f(k) = 1 + 2k$ for integers $0 \le k < N$:

\begin{equation*}
    \sum_{k=0}^{N-1} w_k f(k)
\end{equation*}

A direct computation shows that $\sum_{k=0}^{N-1}w_k f(k) = 36.0$

Using the canonical computation above and the fact that $\sum_{k=0}^{N-1}w_k = \frac{N}{2}$ from
Equation~\ref{eqn:sum_sin2} we obtain  $\sum_{k=0}^{N-1}w_k (1 + 2k) \approx 4 + 2 \cdot 15.99768 = 35.99536$.

\subsection{\label{subsec:value_counting}Value Counting Using Generalized Inner Product}

\begin{figure*}[ht]
    \centering
    \includegraphics[width=.9\linewidth]{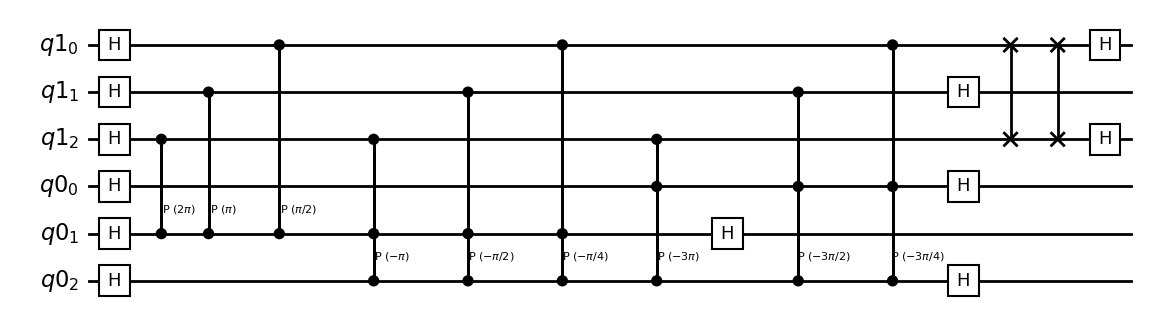}
    \caption{The quantum circuit visualizing the implementation of the example in Section~\ref{subsec:value_counting}
    . The operators $F$ and $B$ both end with the inverse Fourier transform, which cancels out and therefore it can
    be omitted.}
    \label{fig:counting_circuit}
\end{figure*}

Given integers $n = 3$ and $m = 3$, $N = 2^n$ and $M = 2^m$, weights $w_k = 1$ for $0 \le k < N$, hashes

\begin{equation*}
    h_v =
    \begin{cases}
        1, & \text{if } v = v_0\\
        0, & \text{otherwise}\\
    \end{cases}
\end{equation*}

for integers $0 \le v < M$, and a function $f(k_0, k_1, k_2) = 2 k_1 - k_0 k_1 - 3 k_0 k_2$ for $(k_0, k_1,
k_2) \in \{0, 1\}^n$ and a value $v_0$, we want the number of inputs $k \in K$ such that $f(k) = v_0$.

\begin{center}
    \begin{tabular}{c|c}
        $k$ & $f(k)$ \\
        \hline
        $0$ & $0$    \\
        $1$ & $0$    \\
        $2$ & $2$    \\
        $3$ & $2$    \\
        $4$ & $0$    \\
        $5$ & $-3$   \\
        $6$ & $1$    \\
        $7$ & $-2$   \\
    \end{tabular}
\end{center}

Therefore, we are interested in calculating the weighted sum of weighted/hashed function values:

\begin{equation*}
    \sum_{k=0}^{N-1} w_k h_{f(k)}
\end{equation*}

Note that we are using the binary representation of keys, i.e. $k = k_0 + 2 k_1 + 4 k_2 \text{ for } 0 \le k < N$.

Let $A$ be an operator that prepares the state $\sum_{k = 0}^{N-1}a_k\ket{k}_n$, with $a = \frac{1}{\sqrt{N}}$ for $0
\le k < N$, and let $B$ be an operator that prepares the state $\ket{v_0}_m$, with $b = 1$ for $0 \le v < M$, and let
$F$ be an operator that encodes the function $f$.

Using the Controlled Weighted Sum pattern~\ref{pattern2}, we can compute the weighted sum:

\begin{equation*}
    \begin{split}
        \sum_{k=0}^{N-1} w_k h_{f(k)} & = \left|f^{-1}(v_0) \}\right| \\
        & = N \bra{0} (I_n \otimes B^{\dagger}) F (A \otimes I_m)\ket{0}_{n+m}.
    \end{split}
\end{equation*}

Running the computation in a simulator yields the value of $\bra{0} (I_n \otimes B^{\dagger}) F (A \otimes I_m)
\ket{0}_{n+m}$ (i.e. the amplitude of $\ket{0}_{n+m}$) as $0.375$, and therefore:

\begin{equation*}
    \left|f^{-1}(0) \}\right| = 8 \cdot 0.375 = 3
\end{equation*}

so the polynomial $f$ has $3$ zeros.

The quantum circuit that implements this computation as described above is shown in Fig.~\ref{fig:counting_circuit}.

%%%%%%%%%%%%%%%%%%%%%%%%%%%%%%%%%%%%%%%%%%%%%%%%%%%%%%%%%%%%%%%%%%%%%%%%%%%
\section{\label{sec:experiments}Experiments on Quantum Hardware}
%%%%%%%%%%%%%%%%%%%%%%%%%%%%%%%%%%%%%%%%%%%%%%%%%%%%%%%%%%%%%%%%%%%%%%%%%%%

The experiments discussed in this section were run on IBM quantum devices powered by
IBM Quantum Falcon Processors. For each experiment we include the best result of at least 10 runs.

As an experimental observation, we noticed a close correlation between the readout error on
a quantum device and the error in inner product computations that does not generally apply to
other computations. This may be an interesting area of research.

\paragraph{Expected value of discrete functions.}

In this experiment we performed the expected value computation discussed in Section~\ref{subsec:expected_value_poly}
on the IBM ibmq{\_}casablanca 7-qubit device with quantum volume 32. Each run of the experiment was
performed with 8192 shots.

The amplitude of $\ket{0}$ in the best experiment result was $\approx 0.17747$, compared to the
simulated value $\approx 0.17835$, as calculated in Section~\ref{subsec:expected_value_poly}.
The readout assignment error at the time of the experiment was 1.93\%.

\paragraph{Expected value of linear functions for a trigonometric distribution.}

In this experiment we performed the expected value computation discussed in Section~\ref{subsubsec:linear_exact_2}
on the IBM ibmq{\_}jakarta 7-qubit device with quantum volume 16. Each run of the experiment was
performed with 8192 shots.

The amplitude of $\ket{0}$ in the best experiment result was $\approx 0.72996$, compared to the
simulated value $\approx 0.77998$, as calculated in Section~\ref{subsubsec:linear_exact_2}.
A comparison between the outcome probability distributions of the experiment and a simulation is shown in
Fig.~\ref{fig:expected_value_ibmq_jakarta}.
The readout assignment error at the time of the experiment was 4.07\%.

\begin{center}
    \begin{tabular}{c}
        \includegraphics[width=0.6\linewidth]{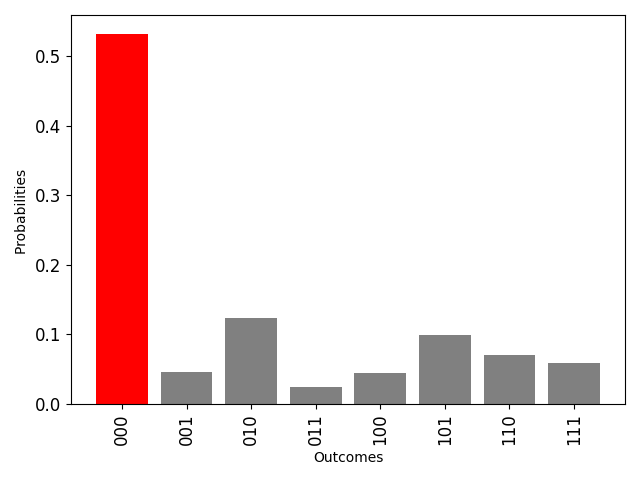} \\
        \includegraphics[width=0.6\linewidth]{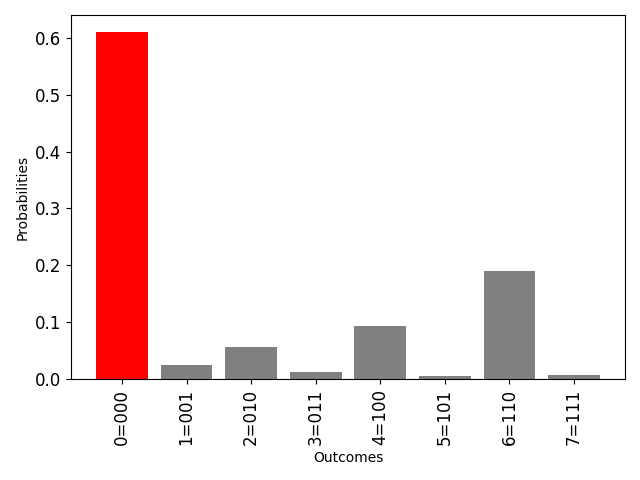} \\
    \end{tabular}
    \captionof{figure}{Visualization of the outcome probabilities of a three-qubit quantum state after
    performing the computation described in Section~\ref{subsubsec:linear_exact_2}, running on ibmq{\_}jakarta (top), and 
    in a simulation environment (bottom). The outcomes different from $\ket{0}$ are grayed out because they do not contribute to the computation result.
    }
    \label{fig:expected_value_ibmq_jakarta}
\end{center}

\paragraph{Expected value of linear functions for a normal distribution.}

In this experiment, we consider the computation of the expected value of the identity function with a normally
distributed argument. Using the Weighted Sum pattern~\ref{pattern1}, where $A$ is the unitary operator that encodes a
normal distribution, as in Appendix~\ref{subsec:catalog_normal_heuristic}, and $B = L_n$  as described in
Equation~\ref{eqn:heuristic_lin}, for $n = 3$ qubits.
The experiment was performed on the IBM ibmq{\_}jakarta 7-qubit device with quantum volume 16.
Each run of the experiment was
performed with 8192 shots.

The amplitude of $\ket{0}$ in the best experiment result was $\approx 0.57378$, compared to the
simulated value $\approx 0.60131$.
The readout assignment error at the time of the experiment was 2.79\%.

%%%%%%%%%%%%%%%%%%%%%%%%%%%%%%%%%%%%%%%%%%%%%%%%%%%%%%%%%%%%%%%%%%%%%%%%%%%
\section{\label{sec:related_work}Related Work}
%%%%%%%%%%%%%%%%%%%%%%%%%%%%%%%%%%%%%%%%%%%%%%%%%%%%%%%%%%%%%%%%%%%%%%%%%%%

Buhrman et al. ~\cite{BurhmanCleve} define a quantum inner product algorithm, the SWAP test, which efficiently solves
the quantum states equality problem with high probability. This algorithm has been used as a building block in several clustering algorithms for
supervised and unsupervised machine learning ~\cite{lloyd2013quantum}.

In the field of quantum deep learning, Tacchino et al. propose an inner product algorithm for implementing quantum
neurons ~\cite{Tacchino2019QuantumNeuron}.

Woerner et al. propose using quantum algorithms for Monte Carlo simulations for financial risk management
~\cite{QuantumRisk}. These algorithms effectively calculate the inner product of a quantum states representing
the distribution of risk factors and asset values.

A widely used class of supervised learning algorithms is Support Vector Machines (SVMs) ~\cite{Liu2020QSSVM}. Their
key computational component is a kernel function that calculates the inner product of feature vectors.
This can be implemented by
using a feature space implemented as a highly dimensional Hilbert space and a quantum inner product kernel
function ~\cite{SchuldQML}.

%%%%%%%%%%%%%%%%%%%%%%%%%%%%%%%%%%%%%%%%%%%%%%%%%%%%%%%%%%%%%%%%%%%%%%%%%%%
\section{\label{sec:conclusions}Concluding Remarks}
%%%%%%%%%%%%%%%%%%%%%%%%%%%%%%%%%%%%%%%%%%%%%%%%%%%%%%%%%%%%%%%%%%%%%%%%%%%

Efficient methods for calculating expected values, variance, value at risk, etc.
are very important in the financial industry.

Such methods rely on computing weighted sums. Finding quantum solutions in
this space is an active area of research.

Most notably, the Woerner-Egger method addresses this family of computations by using
 $Y$-rotations to perform multiplication of amplitudes and
measurement based addition for implementing weighted sums. In particular,
the implementation of the expected value of a canonical linear function
allows the calculation of the expected value of any linear function.

The Generalized Inner Product method introduced in this paper allows the
calculation of the expected value of any discrete function, not necessarily
linear, based on the encoding of a canonical linear function.

It also provides a canonical way to implement input function selection,
and thus calculating value at risk (VaR), and a canonical way to implement
output value selection, which enables straightforward implementations of
comparators needed in option pricing ~\cite{QuantumRisk, StamatopoulosOptionPricing}, and in general concentration
metrics like counting the number of negative values a function takes, or the
number of zeros of a function.

The paper provides additional building blocks like exact and approximate encodings for common statistical
distributions (e.g. the raised cosine and normal distributions) and functions (e.g. linear functions).
Examples of using exact or approximate encoding, and simple or generalized inner product
are also included in the paper.

Even stronger generalizations of quantum inner products can be obtained by replacing unitary operators
$F$ and $H^{\otimes n}$ in Equation~\ref{eqn:inner_prod_f_b}
with more general operators. This is an interesting future research area.

%%%%%%%%%%%%%%%%%%%%%%%%%%%%%%%%%%%%%%%%%%%%%%%%%%%%%%%%%%%%%%%%%%%%%%%%%%%
\acknowledgements
%%%%%%%%%%%%%%%%%%%%%%%%%%%%%%%%%%%%%%%%%%%%%%%%%%%%%%%%%%%%%%%%%%%%%%%%%%%

The authors want to thank Vitaliy Dorum for helping with the development of this manuscript.\\

The views expressed in this article are those of the authors and do not represent the views of Wells Fargo. This article
is for informational purposes only. Nothing contained in this article should be construed as investment advice.
Wells Fargo makes no express or implied warranties and expressly disclaims all legal, tax, and accounting implications
related to this article.\\

We acknowledge the use of IBM Quantum services for this work.
The views expressed are those of the authors, and do not
reflect the official policy or position of IBM or the IBM Quantum team.

%%%%%%%%%%%%%%%%%%%%%%%%%%%%%%%%%%%%%%%%%%%%%%%%%%%%%%%%%%%%%%%%%%%%%%%%%%%s
\bibliographystyle{unsrtnat}
\bibliography{main}% Produces the bibliography via BibTeX.
%%%%%%%%%%%%%%%%%%%%%%%%%%%%%%%%%%%%%%%%%%%%%%%%%%%%%%%%%%%%%%%%%%%%%%%%%%%

%%%%%%%%%%%%%%%%%%%%%%%%%%%%%%%%%%%%%%%%%%%%%%%%%%%%%%%%%%%%%%%%%%%%%%%%%%%
\onecolumn\newpage
\renewcommand\floatpagefraction{0.9}
\appendix
%%%%%%%%%%%%%%%%%%%%%%%%%%%%%%%%%%%%%%%%%%%%%%%%%%%%%%%%%%%%%%%%%%%%%%%%%%%

%%%%%%%%%%%%%%%%%%%%%%%%%%%%%%%%%%%%%%%%%%%%%%%%%%%%%%%%%%%%%%%%%%%%%%%%%%%
\section{\label{sec:linear_sw}The Woerner-Egger Approximation Method}
%%%%%%%%%%%%%%%%%%%%%%%%%%%%%%%%%%%%%%%%%%%%%%%%%%%%%%%%%%%%%%%%%%%%%%%%%%%

\begin{theorem-non3}
    \label{woerner_egger}
    Given an integer $n > 0$ and $N = 2^n$, probabilities $p_k \in [0, 1]$ defined for integers $0 \le k < N$,
    with $\sum_{k=0}^{N-1} p_k = 1$, and a
    function $f: \{0, \mathellipsis, N-1\} \rightarrow [-1, 1]$ we are interested in calculating the sum of products

    \begin{equation*}
        \label{eqn:woerner_egger}
        \sum_{k=0}^{N-1}p_k f(k).
    \end{equation*}
\end{theorem-non3}

\begin{proof}[Solution Sketch]
    We paraphrase~\cite{QuantumRisk} in this description. We can use the approximation

    \begin{equation*}
        y \approx \frac{1}{2}\sin(2y) = \sin^2(y + \frac{\pi}{4}) - \frac{1}{2} \text{ for } y \approx 0.
    \end{equation*}

    Then for a constant $c \approx 0$ we have

    \begin{equation*}
        \begin{split}
            \sum_{k=0}^{N-1}p_k f(k) & \approx \sum_{k=0}^{N-1}p_k \frac{1}{2c}\sin(2cf(k)) = \sum_{k=0}^{N-1}p_k
            \frac{1}{c} \left(\sin^2\left(cf(k) + \frac{\pi}{4}\right) - \frac{1}{2}\right)\\
            & = \frac{1}{c}\sum_{k=0}^{N-1}p_k \left(\sin^2\left(cf(k) + \frac{\pi}{4}\right)\right) - \frac{1}{2c}
            = \frac{1}{c}\sum_{k=0}^{N-1}\left(\sqrt{p_k}\sin\left(cf(k) + \frac{\pi}{4}\right)\right)^2 -
            \frac{1}{2c}
        \end{split}
    \end{equation*}

    The products $\sqrt{p_k}\sin(\theta_k)$ for $\theta_k \in \mathbb{R}$ can be encoded as amplitudes of
    a quantum states using $R_Y$ single-qubit quantum gates acting on an ancillary qubit as described
    in~\cite{QuantumRisk}. In essence, given a quantum state that encodes the probability distribution $p$

    \begin{equation*}
        \sum_{k=0}^{N-1}\sqrt{p_k} \ket{k}_n
    \end{equation*}

    the following quantum state can be built by adding an ancillary qubit and applying (controlled) $R_Y$ gates
    to it

    \begin{equation*}
        \sum_{k=0}^{N-1}\sqrt{p_k} \cos(\theta_k)\ket{k}_n\ket{0} + \sum_{k=0}^{N-1}\sqrt{p_k}  \sin(\theta_k)
        \ket{k}_n\ket{1},
    \end{equation*}

    where $\theta_k = cf(k) + \frac{\pi}{4}$.

    The probability of measuring $\ket{1}$ in the ancillary qubit is

    \begin{equation*}
        P_1 = \sum_{k=0}^{N-1}p_k  \sin^2(\theta_k) \approx \sin^2(\frac{\pi}{4}) \sum_{k=0}^{N-1}p_k = \frac{1}{2},
    \end{equation*}

    and it can be approximated using amplitude estimation algorithms.Then

    \begin{equation*}
        \sum_{k=0}^{N-1}p_k f(k) \approx \frac{1}{2c}(2P_1 - 1).
    \end{equation*}
\end{proof}

Note that the summation is done at the probability level, through measurement, not in amplitudes.

The particular case of linear functions is treated in~\cite{StamatopoulosOptionPricing} where it is also
pointed out that considering only the canonical linear function $f(k) = -1 + \frac{2}{N-1} k$ is sufficient.

Note that:

\begin{equation*}
    -1 \le -1 + k\frac{2}{N-1} \le 1 \text{ and } \frac{1}{2} - c \le c\left(-1 + k\frac{2}{N-1}\right)+\frac{1}{2}
    \le \frac{1}{2} + c.\\
\end{equation*}

Adding up the probabilities for outcomes with the ancillary qubit measured as $\ket{1}$, we get:

\begin{equation*}
    \begin{split}
    P_1 & = \sum_{k=0}^{N-1}p_k \sin^2\left(c(-1 + k\frac{2}{N-1}) + \frac{\pi}{4}\right) \approx \sum_{k=0}^{N-1}p_k
    \left(c(-1 + k\frac{2}{N-1}) + \frac{1}{2}\right) \\
    & = \left(\frac{1}{2} - c\right)\sum_{k=0}^{N-1}p_k + c\frac{2}{N-1}\sum_{k=0}^{N-1}p_k k = \frac{1}{2} - c +
    c\frac{2}{N-1}\sum_{k=0}^{N-1}p_k k
    \end{split}
\end{equation*}

Therefore

\begin{equation*}
     \sum_{k=0}^{N-1} p_k k \approx \frac{N-1}{2c} \left( P_1 - \frac{1}{2} + c\right).
\end{equation*}

This can be used to calculate $\sum_{k=0}^{N-1} p_k f(k)$ for any linear function $f$ by using its intercept
and slope. Alternatively, as in~\cite{StamatopoulosOptionPricing}, we can use the fact that a linear function
is completely determined by its minimum and maximum, to obtain

\begin{equation*}
    \sum_{k=0}^{N-1} p_k f(k) \approx m + \frac{M - m}{2c}  \left( P_1 - \frac{1}{2} + c\right).
\end{equation*}

for a linear function $f: \{0, \mathellipsis, N-1\} \rightarrow [m, M]$.

\begin{figure}
    \centering
    \includegraphics[width=.4\textwidth]{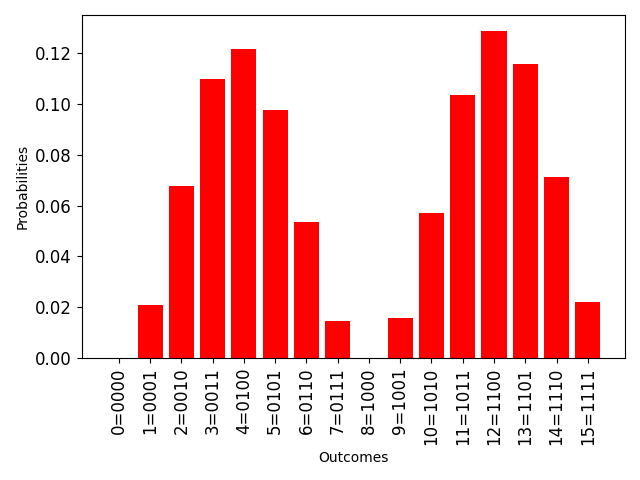}
    \caption{Outcome probability distribution of a quantum state that uses three main qubits and an ancillary qubit,
        prepared using the \textbf{Sum of Products} method.}
    \label{fig:multiplication_sw_probs}
\end{figure}

Intuitively, around half (between $\frac{1}{2} - c$ and $\frac{1}{2} + c$) of each original amplitude
is moved to the corresponding one that measures $\ket{1}$ in the ancilla.

%%%%%%%%%%%%%%%%%%%%%%%%%%%%%%%%%%%%%%%%%%%%%%%%%%%%%%%%%%%%%%%%%%%%%%%%%%%
\section{\label{sec:catalog}Heuristic Circuit Catalog}
%%%%%%%%%%%%%%%%%%%%%%%%%%%%%%%%%%%%%%%%%%%%%%%%%%%%%%%%%%%%%%%%%%%%%%%%%%%

\subsection{\label{subsec:catalog_linear_heuristic}Encoding of the Identity Function}

The following heuristic circuits provide implementations of a linear function, as discussed in
Section~\ref{subsec:identity}.

\begin{center}
    \begin{tabular}{cc}
        \begin{minipage}{.4\textwidth}
            \centering
            \includegraphics[width=.7\linewidth]{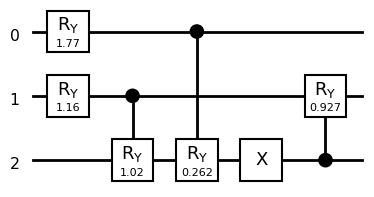}
        \end{minipage}
        \begin{minipage}{.4\textwidth}
            \centering
            \includegraphics[width=.8\linewidth]{identity_q3.png}
        \end{minipage}
    \end{tabular}
    \captionof{figure}{Left: The quantum circuit that encodes the quantum state $\ket{\lambda}_3$ using $n = 3$ qubits, as
    defined in Eq .~\ref{eqn:heuristic_lin}. Right: Visualization of the amplitudes of the quantum state with $n =
    3$ qubits, prepared using the circuit in the figure.}
    \label{fig:identity_q3_circuit}
\end{center}

\begin{center}
    \begin{tabular}{cc}
        \begin{minipage}{.4\textwidth}
            \centering
            \includegraphics[width=.8\linewidth]{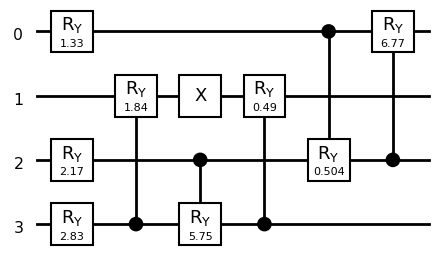}
        \end{minipage}
        \begin{minipage}{.4\textwidth}
            \centering
            \includegraphics[width=.8\linewidth]{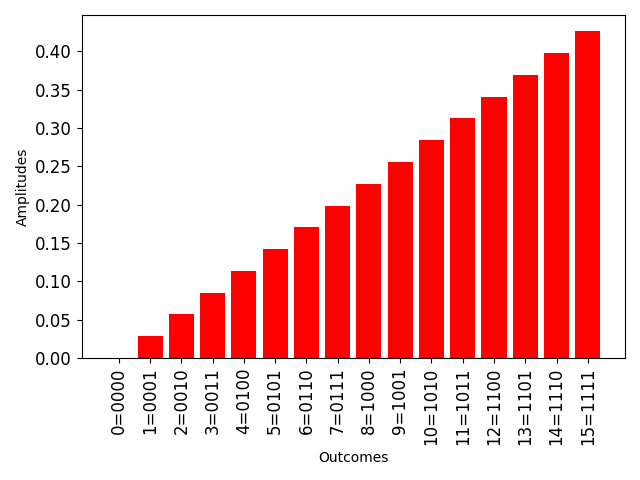}
        \end{minipage}
    \end{tabular}
    \captionof{figure}{Left: The quantum circuit that encodes the quantum state $\ket{\lambda}_4$ using $n = 4$
        qubits, as defined in Eq .~\ref{eqn:heuristic_lin}. Right: Visualization of the amplitudes of a quantum
        state with $n = 4$ qubits, prepared using the circuit in the figure.}
    \label{fig:identity_q4_circuit}
\end{center}

\begin{center}
    \begin{tabular}{cc}
        \begin{minipage}{.4\textwidth}
            \centering
            \includegraphics[width=.8\linewidth]{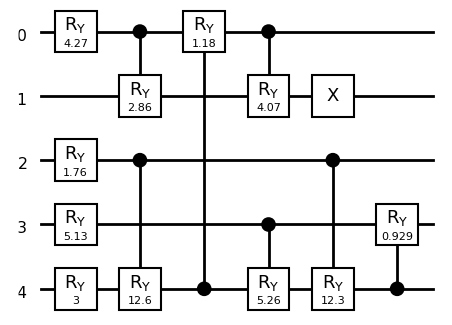}
        \end{minipage}
        \begin{minipage}{.4\textwidth}
            \centering
            \includegraphics[width=.8\linewidth]{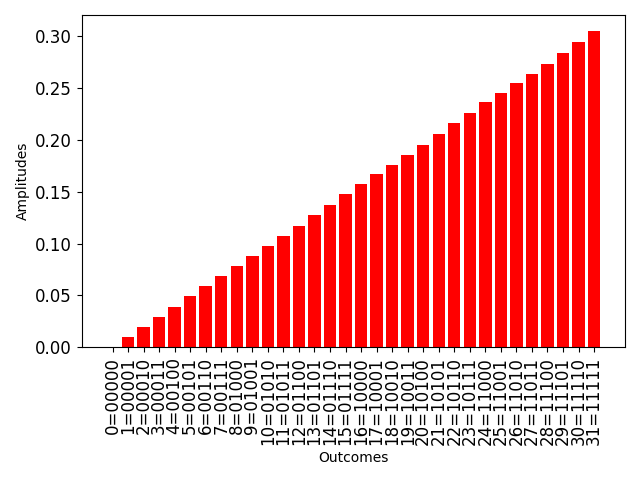}
        \end{minipage}
    \end{tabular}
    \captionof{figure}{Left: The quantum circuit that encodes the quantum state $\ket{\lambda}_5$ using $n = 5$
        qubits, as defined in Eq .~\ref{eqn:heuristic_lin}. Right: Visualization of the amplitudes of a quantum
        state with $n = 5$ qubits, prepared using the circuit in the figure.}
    \label{fig:identity_q5_circuit}
\end{center}

\subsection{\label{subsec:catalog_normal_heuristic}Encoding a Normal Distribution}

The following heuristic circuits encode a normal distribution in a quantum state with three or four qubits,
respectively.

\begin{center}
    \begin{tabular}{cc}
        \begin{minipage}{.4\textwidth}
            \centering
            \includegraphics[width=.8\linewidth]{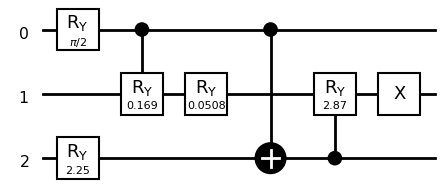}
        \end{minipage}
        \begin{minipage}{.4\textwidth}
            \centering
            \includegraphics[width=.8\linewidth]{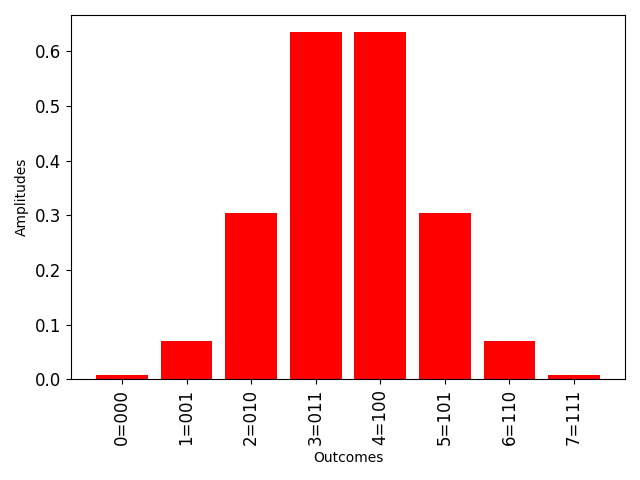}
        \end{minipage}
    \end{tabular}
    \captionof{figure}{Left: The quantum circuit that encodes a normal distribution in a quantum state with three
    qubits. Right: Visualization of the amplitudes of a three-quantum state prepared using the circuit in the figure.}
    \label{fig:normal_heuristic_q3_circuit}
\end{center}

\begin{center}
    \begin{tabular}{cc}
        \begin{minipage}{.45\textwidth}
            \centering
            \includegraphics[width=.8\linewidth]{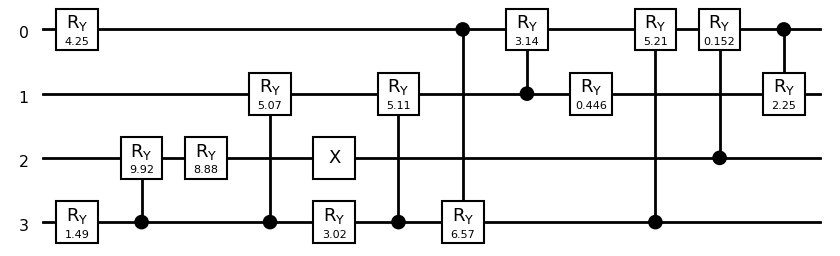}
        \end{minipage}
        \begin{minipage}{.4\textwidth}
            \centering
            \includegraphics[width=.8\linewidth]{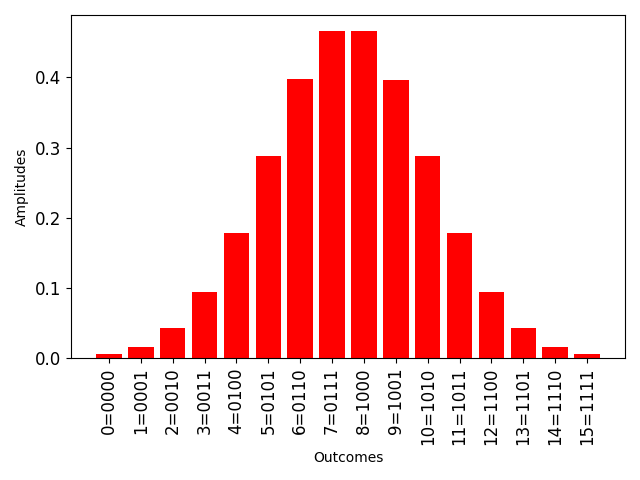}
        \end{minipage}
    \end{tabular}
    \captionof{figure}{Left: The quantum circuit that encodes a normal distribution in a quantum state with four
    qubits. Right: Visualization of the amplitudes of a four-quantum state prepared using the circuit in the figure.}
    \label{fig:normal_heuristic_q4_circuit}
\end{center}

\end{document}